%% file: rt_bd_manuscript_vikram.tex
\documentclass[%
aip,
amsmath,amssymb,
preprint,%
floatfix,
]{revtex4-1}
\usepackage{graphicx}
\usepackage{dcolumn}
\usepackage{bm}
\usepackage{placeins}
\usepackage[utf8]{inputenc}
\usepackage[T1]{fontenc}
\usepackage{mathptmx}
\usepackage{xcolor}
\usepackage{sidecap}
\sidecaptionvpos{figure}{c} 
\begin{document}
	\title[]{A numerical study of gravity driven instability in strongly coupled dusty plasmas. Part I: Rayleigh-Taylor  instability and Buoyancy-driven instability}
	\author{Vikram S. Dharodi}%
	\email{dharodiv@msu.edu}
	\affiliation{Mechanical Engineering, Michigan State University,
		East Lansing, Michigan 48824, USA.}
	
	\author{Amita Das}%
	\email{amita@iitd.ac.in}
	\affiliation{Department of Physics, Indian Institute of Technology, New Delhi, 110016, Delhi, India}%
	
	
	\date{\today}
	
	\begin{abstract}
		Rayleigh-Taylor and Buoyancy-driven instabilities are very common instabilities for an inhomogeneous medium. We examine here how these instabilities grow for incompressible viscoelastic fluids like a strongly coupled dusty plasma by using incompressible generalized hydrodynamic (i-GHD) fluid model. Since the dust particles are pretty massive, the gravitational attraction of earth has a significant role in its dynamics. In this paper the acceleration due to gravity g and the role of strong coupling in the context of gravitationally stratified dusty plasma fluid have been considered. We find that the appearance of elasticity speed up the growth of viscoelastic RT instability by reducing the effect of viscosity at timescales shorter than the Maxwell relaxation time $\tau_m$. The buoyancy driven situation with spatially localized (in both dimensions) of low/high density regions placed in a background medium has also been evolved in the presence of gravity.  Our studies show that both the instabilities get suppressed  with increasing coupling strength of the medium. This suppression has been illustrated analytically as well as by carrying out a two-dimensional fully nonlinear simulation. 
	\end{abstract}
	
	\maketitle
	\section{Introduction}\label{Introduction}
	
	The Rayleigh-Taylor (RT) instability is one of the most prominent fluid instabilities. This instability, for a fluid in a gravitational field, was first observed by Rayleigh  \cite{rayleigh1900investigation} and later implemented to all accelerated fluids by Taylor \cite{taylor1950instability}. Since then numerous experiments and numerical simulations have been conducted in order to better understand it. For a hydrodynamic (HD) fluid under gravity, a detailed theoretical work has been done by S. Chandrasekhar \cite{chandrhd}. A  lot of research studies has been pursued and published on this topic  so  far \cite{youngs1984numerical,waddell2001experimental,livescu2004comp,guo2010linear,baldwin2015inhibition,lyubimova2019rayleigh}. Furthermore, in the context of plasma medium the charged species respond to electromagnetic forces and various scenarios arise where an inverse stratification against a force (and/or a pseudo force due to the choice of the frame) leads to this instability. Studies  for different types of plasmas under various conditions have been carried out \cite{dickinson1962experimental,ariel1970rayleigh,takabe1985self,mikhailenko2002rayleigh,ma2006behavior,sen2010rayleigh, weber2014inhibition,kessler2014gravitational, haines2014plasma,hoshoudy2014rayleigh,khiar2019laser}. This instability has been observed in diverse situations such as a  supernova explosion \cite{Norman_Supernova,Arnett_Supernova}, geophysics \cite{plag1995rayleigh}, astrophysics \cite{1984MNRAS, Zingale_2005}, atmospheric physics\cite{keskinen1981nonlinear}, liquid atomization \cite{Reitz_1999,kong1999developments}, fusion physics \cite{Chertkov2003,Kadau2004,hoshoudy2011quantum,Srinivasan2013}, oceanography \cite{debnath1994nonlinear}, turbulent mixing \cite{Abarzhi1,Abarzhi2,Abarzhi3}, etc. Another form of RT instability is the buoyancy-driven (BD) instability, here a buoyant force acts in the direction opposite to gravity for low density regions. The BD instability decides whether an object will float (if the density of an object is less than background fluid) or sink (if the density of an object is greater than background fluid) in fluid. The floating of boats and ships while sinking of small objects like rocks in water, or the pouring cream (heavy fluid) into coffee (light fluid) and petroleum wells are some examples where buoyant force plays a major role.
	
	Our objective here is to understand the both RT  and BD instabilities  when a medium is in strongly coupled state. For this purpose we consider the case of a dusty plasma. A dusty plasma can be  easily prepared/found in the strong coupling state  and also the gravity plays an important role for a micron-sized dust particle has a mass of about  $10^{-15}-10^{-10}$Kg. Thus the role of strong coupling in behaviour of  these instabilities can be readily investigated in the context of strongly coupled dusty plasmas (SCDPs). The SCDPs are modelled, here, by using the incompressible generalized hydrodynamic (i-GHD) fluid model.  This model captures the strong coupling behaviour in terms of a viscoelastic fluid characteristics through two crucial parameters, the  shear viscosity $\eta$ and elasticity $\tau_m$ (representing memory relaxation time) \cite{frenkel_kinetic}.  For those phenomena which are faster compared to $\tau_m$, the GHD system retains the memory of the past configurations and the elasticity effects dominate.  This elastic behaviour of the system is known to produce transverse shear (TS) waves in the medium. However, for times longer than $\tau_m$ the memory effects are insignificant and the usual viscous characteristics of the fluid phase dominate.
	
	Here, we demonstrate that elasticity leads to a speed up of the RT instability by reducing the coupling strength ($\eta$/$\tau_m$, viscous fraction). Such reducing effect of elasticity on RT instability has also been observed for geodynamic settings in \cite{kaus2007effects}  and for the viscoelastic fluids within the framework of the Oldroyd-B model in~\cite{boffetta_mazzino_musacchio_vozella_2010}.  In order to evident this effect, for the same value of $\eta$, we have simulated the fully viscous fluid and viscoelastic fluids for different values of $\tau_m$. To develop a  better physical insight into the dynamics of each phenomena, the inviscid limit of fluids are also simulated. The BD instability has been discussed by considering the evolution of localized low/high density bubbles/droplets in the medium. For both instabilities, the simulations show a gradual suppression with increasing coupling strength, which is consistent with the linear theory.  The coupling strength has been understood in terms of shear wave whose phase velocity is proportional to it.
	
	The paper has been organized as follows. Section \ref{anal_description} contains the governing equations of i-GHD model for viscoelastic medium in the presence of gravity. In Subsections \ref{Sub_anal_dispertion_GD} and \ref{Sub_anal_dispertion_SD}, we obtain the analytical dispersion relations for the gradually and sharply increasing density gradients against gravitational force, respectively. Section \ref{num_methodology} contains the description of numerical scheme.  In Section \ref{num_gd_bd}, we describe  the numerical evolution of various density profiles with time for different values of coupling parameters. The suppression of the both instabilities i.e. RT and BD has been clearly depicted by numerical simulation, as one moves from weakly coupled to strongly coupled regime. In Section \ref{conclusions}, we describe the results and observations.
	
	\section{Analytical Description} \label{anal_description}
	
	Generalized hydrodynamic (GHD) fluid model is a phenomenological model which is used to study the strongly coupled dusty plasma system \cite{Kaw_Sen_1998, Kaw_2001, tiwari2012kelvin, dharodi2014visco, tiwari2014evolution, dharodi2016sub}. GHD model treats dusty plasma as a viscoelastic fluid in which the coupling strength depends on the ratio of the  shear viscosity coefficient $\eta$ and relaxation time parameter $\tau_m$~\cite{frenkel_kinetic}. This model supports both the existence of incompressible transverse shear and compressible longitudinal modes. Here, we consider only the  incompressible limit of the GHD  and term is as the i-GHD fluid model for which the  set of equations is obtained. In the incompressible limit the Poisson equation is replaced by the quasi-neutrality condition and charge density fluctuations are ignored. The coupled set of continuity and momentum equations for the dust fluid under gravity acceleration $\vec{g}$ can be written as:
	\begin{equation}\label{eq:continuity}
	\frac{\partial \rho_d}{\partial t} + \nabla \cdot
	\left(\rho_d\vec{v}_d\right)=0{,}
	\end{equation}
	\begin{eqnarray}\label{eq:momentum}
	&&\left[1+{\tau_m}\left(\frac{\partial}{\partial{t}}+{\vec{v}_d}\cdot \nabla\right)\right]\nonumber\\
	&& \left[ {{\rho_d}\left(\frac{\partial{\vec{v}_d}}{\partial {t}}+{\vec{v}_d}{\cdot} \nabla{\vec{v}_d}\right)}+{\rho_d}\vec{g}+{\rho_c}\nabla \phi_{d} \right]\nonumber\\
	&& =\eta \nabla^2\vec{v}_d{,}
	\end{eqnarray}
	respectively and the incompressible condition is given as
	\begin{equation}\label{eq:incompressible}
	{\nabla}{\cdot}{\vec{v_d}}=0{.}
	\end{equation}
	The derivation of these reduced equations has been discussed in detail in our earlier papers \cite{dharodi2014visco,dharodi2016sub} along with the procedure of its  numerical implementation and validation.  Here, the variables $\rho_d$ and $\rho_c$, $\vec{v}_d$ and $\phi_d$ are the   dust mass density, dust charge density, dust fluid velocity and dust charge potential respectively. The time, length, velocity and potential are normalised by inverse of dust plasma frequency $\omega^{-1}_{pd} = \left({4\pi (Z_d e)^{2}n_{d0}}/{m_{d0}}\right)^{-1/2}$ and plasma Debye length $\lambda_{d} = \left({K_B T_i}/{4{\pi} {Z_d}{n_{d0}}{e^2}}\right)^{1/2}$, ${\lambda_d}{\omega_{pd}}$ and ${{Z_d}e}/{{K_B}{T_i}}$ respectively. The parameters $m_d$, $T_i$ and $K_B$ are the dust grain mass, ion temperature and Boltzmann constant respectively. $Z_d$ is the charge on each dust grain with no consideration  of charge fluctuation. The number density $n_d$ is normalised by the equilibrium value $n_{d0}$. 
	\subsection{Gradual density gradient} \label{Sub_anal_dispertion_GD}
	
	Here, we consider two-dimensional (x-y plane) incompressible system where density gradient and potential gradient are chosen along the y-axis i.e. ${{\partial{\rho_d}}/{\partial y}}$, ${{\partial \phi_{d0}}/{\partial y}}$ respectively, and acceleration $\vec g$ applied in opposite direction of fluid density gradient $-g\hat{y}$. Initially, we consider no initial flow i.e. $\vec v_{d0}=0$ at t=0, the equilibrium condition Eq.~(\ref{eq:momentum}) becomes
	\begin{equation}\label{eq:equilibrium}
	{\rho_{d0}}{g} =-{\rho_c}{\frac{\partial\phi_{d0}} {\partial y}}{.}
	\end{equation}
	A small perturbation in the various fields, e.g. density,  scalar potential and dust velocity can be written as 
	\begin{equation}\label{eq:pert}
	{\rho_d}(x,y,t)={\rho_{d0}}(y,t=0)+{\rho_{d1}}(x,y,t) {,}
	\end{equation}
	\begin{equation}
	{\phi_d}(x,y,t)={\phi_{d0}}(y,t=0)+{{\phi_{d1}}}(x,y,t){,}
	\end{equation}
	\begin{equation}
	{\vec{v}_d}(x,y,t)=0+{\vec{v}_{d1}}(x,y,t) {,}
	\end{equation}
	respectively. Retaining only linear terms in the perturbed fields we obtain the following equations for the linearized instability analysis
	\begin{equation}
	\frac{\partial \rho_{d1} }{\partial t} + \left(\vec{v_{d1}}\cdot
	\nabla\right)\rho_{d0}= 0{,}
	\end{equation}
	\begin{equation}
	\left[1 + \tau_m \frac{\partial} {\partial t}\right]
	\left[{{\rho_{d0}}{\frac{\partial {v}_{d1y}}  {\partial
				t}}}+{\rho_{d1}}g{\hat{y}}
	+{\rho_c}{\frac{\partial \phi_{d1y}}{\partial y}}\right]=\eta \nabla^2
	{v}_{d1y}{,}   
	\end{equation}
	\begin{equation}
	\left[1 + \tau_m \frac{\partial} {\partial t}\right]
	\left[{{\rho_{d0}}{\frac{\partial {v}_{d1x}}  {\partial t}}}
	+{\rho_c}{\frac{\partial \phi_{d1x}}{\partial x}}\right]=\eta \nabla^2
	{v}_{d1x}{,}   
	\end{equation}
	\begin{equation}
	\nabla \cdot \vec v_{d1}=0{.}
	\end{equation}
	Since equilibrium fields vary along $\hat{y}$, the above set of equations can  only be Fourier analysed in time and spatial coordinate $x$. However, assuming the perturbations to be of much smaller scale compared to the equilibrium  variation along $y$ we invoke the local approximation and proceed with the Fourier decomposition along y also. This leads to 
	\begin{equation}\label{eq:di}
	{\rho_{d1} }=-\frac{i}{{\omega}}\frac{\partial \rho_{d0} }{\partial
		y}{v_{1y}}{.}  
	\end{equation}
	The $\hat{y}$ component 
	\begin{equation}\label{eq:v1y}
	{\left[1-i{\omega}\tau_{m}\right]}
	{\left[{-i{\omega}{\rho_{d0}}{{v}_{1y}}}
		+{\rho_{d1}}g+i{\mathrm{k}_y}{\rho_c}{\phi_1}\right]}=-{\eta
		{\mathrm{k}}^2{v}_{1y}}{.}
	\end{equation}
	The $\hat{x}$ component
	\begin{equation}\label{eq:v1x}
	{\left[1-i{\omega}\tau_{m}\right]}{\left[{-i{\omega}{\rho_{d0}}{{v}_{1x}}}+i{
			\mathrm{k}_x}{\rho_c}{\phi_1}\right]} =-{\eta {\mathrm{k}}^2
		{v}_{1x}}{,} 
	\end{equation}
	\begin{equation}\label{eq:incom_fft}
	i{\mathrm{k}_x}{v_{1x}}=-i{\mathrm{k}_y}{v_{1y}}
	\Rightarrow {v_{1y}}={-\frac{\mathrm{k}_x}{\mathrm{k}_y}}{v_{1x}}{.}
	\end{equation}
	Obtain $\phi_{d1}$, using above relation and Eqs.~(\ref{eq:di})~(\ref{eq:v1y})~(\ref{eq:v1x}), and (\ref{eq:incom_fft}), we get 
	\begin{equation}\label{eq:phi1}
	{{\phi_{d1}}}
	=-{\frac{{g}{\mathrm{k}_x}{v_{1x}}}{{\rho_c}{\omega}{\mathrm{k}^2}}}
	\frac{\partial \rho_{d0} }{\partial y}{,}
	\end{equation}
	using the above relation Eq.~(\ref{eq:phi1}) in Eq.~(\ref{eq:v1x}), we get the  dispersion relation as
	\begin{equation}\label{eq:general_disperssion}
	{\left(1-i{\omega}\tau_{m}\right)}{\left[{\omega^2}{\mathrm{k}^2}+
		{\frac{g\mathrm{k}^2_x}{{{\rho_{d0}}}}}
		\frac{\partial \rho_{d0} }{\partial
			y}\right]}={-i{\omega}{\eta^*}{\mathrm{k}}^4}
	{,} 
	\end{equation}
	where kinematic viscosity  $\eta^*$=${\eta}/{\rho_{d0}}$,  the ratio of absolute viscosity ${\eta}$ to density ${\rho_{d0}}$. 
	A similar equation for the electron ion plasma in the context of inertial fusion has been predicted by A. Das {\it et. al.}  \cite{Das_2014}. When $\tau_m$ the memory relaxation parameter is zero, 
	the RT growth rate shows a viscous damping due to $\eta^{*}$. 
	In strongly coupled regime i.e. ${\omega}{\tau_m}\gg 1$, the above dispersion Eq.~(\ref{eq:general_disperssion}) becomes 
	\begin{equation} \label{eq:strong_smooth_dispersion}
	{\omega^2}={\frac{\eta^*}{{\tau_{m}}}{\mathrm{k}}^2}-{\frac{g}{{{\rho_{d0}}}}}
	\frac{\partial \rho_{d0} }{\partial
		y}{\frac{{\mathrm{k^2_x}}}{{\mathrm{k}}^2}}{.}
	\end{equation}
	Here $\sqrt{\eta^{*}/\tau_m}= V_s$ is the phase speed of the transverse shear wave. For hydrodynamic fluid case where ($\eta^*/\tau_m$=0), the above dispersion relation reduces to the usual expression of the RT growth rate is
	\begin{equation} \label{eq:fluid_smooth_dispersion}
	{\omega^2}=-{\frac{g}{{{\rho_{d0}}}}}
	\frac{\partial \rho_{d0} }{\partial
		y}{\frac{{\mathrm{k^2_x}}}{{\mathrm{k}}^2}}{.}
	\end{equation}
	
	\subsection{Sharp interface}
	\label{Sub_anal_dispertion_SD}
	
	When the density interface is sharp the above dispersion relation  becomes,
	\begin{equation}\label{eq:strong_sharp_dispersion}
	{\omega^2}={\frac{\eta^*}{{\tau_{m}}}{{\mathrm{k}}^2}}-g{\mathrm{k}}{A_T}{,}
	\end{equation}
	here $A_T={(\rho_{d02}-\rho_{d01})}/{(\rho_{d01}+\rho_{d02})}$ is the Atwood's number.  $\rho_{d02}$ is heavier fluid density is supported by lighter fluid density $\rho_{d01}$. Here, too the hydrodynamic limit is recovered when the TS wave velocity $\eta^{*}/\tau_m = 0$ is chosen. 
	\begin{equation} \label{eq:fluid_sharp_dispersion}
	{\omega^2}=-g{\mathrm{k}}{A_T}{.}
	\end{equation}
	For $g=0$, Eqs.~(\ref{eq:strong_smooth_dispersion}) and (\ref{eq:strong_sharp_dispersion}) imply the existence of TS wave moving with phase velocity ${\sqrt{{\eta^*}/{\tau_m}}}$. Detailed analytical and numerical studies of  TS waves for  the homogeneous ~\cite{dharodi2014visco,dharodi2016sub} and inhomogeneous~\cite{dharodi2020rotating} density media have been carried out by us. From these two equations one can immediately infer that  in the strong coupling limit as  the value of the first term ${{\eta^*}/{\tau_{m}}}$ (coupling strength parameter) increases, it suppresses  the RT instability growth rate. Alternatively,  this suppressing effect of  strong coupling  on RT instability can also  be understood in terms of shear waves whose phase velocity (${\sqrt{{\eta}/{{\rho_d}{\tau_m}}}}$) is proportional to it. These equations also imply that elasticity, in the viscoelastic fluids, enhances the growth of RT instability by reducing the coupling strength . The effects of elasticity on the RT instability have been given in a number of studies ~\cite{biot1965mechanics,biot1965theory, ode1966gravitational, poliakov1993explicit,naimark1994gravitational,kaus2007effects,boffetta_mazzino_musacchio_vozella_2010}. Most of these studies showed that elasticity speedup the growth of the RT instability. It should be noted that the analytical dispersion relations for dusty plasma medium has also been obtained in 2015 by Avinash {\it et. al.}\cite{avinash2015rayleigh}. However,  we had presented this result earlier in  2014 in a conference\cite{rt_dharodi2014}. 
	
	\section{Numerical simulation} \label{num_methodology}
	
	For the purpose of numerical simulation the generalized momentum Eq.~(\ref{eq:momentum}) has been expressed as a set of following two coupled convective equations,
	\begin{eqnarray}\label{eq:vort_incomp1}
	{{\rho_d}\left(\frac{\partial{\vec{v}_d}}{\partial {t}}+{\vec{v}_d}{\cdot} \nabla{\vec{v}_d}\right)}+{\rho_d}\vec{g}+{\rho_c}\nabla \phi_{d} ={\vec \psi}
	\end{eqnarray}
	\begin{equation}\label{eq:psi_incomp1}
	\frac{\partial {\vec \psi}} {\partial t}+\vec{v}_d \cdot \nabla{\vec \psi}=
	{\frac{\eta}{\tau_m}}{\nabla^2}{\vec{v}_d }-{\frac{\vec \psi}{\tau_m}}{.}
	\end{equation}
	For  two-dimensional (2D) studies all the  variables are functions of  $x$ and $y$ only,  $i.e$ ${\vec \psi}(x,y)$, ${\vec v_d}(x,y)$, ${\rho_d}(x,y)$ and $\vec{g}(y)$. From Eq.~(\ref{eq:vort_incomp1}) it is clear that the quantity ${\vec \psi}(x,y)$ is the strain created in the elastic medium by the time-varying velocity fields. We re-write the Eq.~(\ref{eq:vort_incomp1}) using the  equilibrium condition of  $ {\rho_{d0}}{g} =-{\rho_c}{\nabla\phi_{d0}}$ in terms of perturbed density Eq.~(\ref{eq:pert}).
	\begin{eqnarray}\label{eq:vort_incomp2}
	\frac{\partial{\vec{v}_d}}{\partial {t}}+{\vec{v}_d}{\cdot} \nabla{\vec{v}_d}+{\frac{\rho_{d1}}{\rho_d}}\vec{g}+{\frac{\rho_{c}}{\rho_d}}{\nabla{\phi_{d1}}} =\frac{\vec \psi}{\rho_d}{.}
	\end{eqnarray}
	Taking the curl of Eq.~(\ref{eq:vort_incomp2}) and using the Boussinesq approximation, we get  
	\begin{equation}\label{eq:vort_incomp3} 
	\frac{\partial{\xi_{z}}} {\partial t}+\left(\vec{v}_d \cdot \vec \nabla\right)
	{\xi_{z}}={\frac{1}{\rho_{d0}}}{{\nabla}{\times}{\rho_{d1}}{\vec{g}}}+{\nabla}{\times}{\frac{\vec \psi}{\rho_d}}{.} 
	\end{equation}
	Here, $ {\xi_{z}}(x,y)=\left[\vec{\nabla}{\times}{\vec v_d}(x,y) \right]_z $ is the $z$ (and the only finite) component of vorticity for the 2D case considered here. It is  normalised by dust plasma frequency. The acceleration $\vec g$ is applied  opposite to the  fluid density gradient $-g\hat{y}$.
	The  coupled set of  model equations for numerical simulations are the following: 
	\begin{equation}\label{eq:cont_incomp3}
	\frac{\partial \rho_d }{\partial t} +  \left(\vec{v}_d\cdot
	\nabla\right)\rho_d= 0{,}
	\end{equation}
	\begin{equation}\label{eq:psi_incomp3}
	\frac{\partial {\vec \psi}} {\partial t}+\left(\vec{v}_d \cdot \vec
	\nabla\right)
	{\vec \psi}={\frac{\eta}{\tau_m}}{\nabla^2}{\vec{v}_d }-{\frac{\vec
			\psi}{\tau_m}}{,}  
	\end{equation}
	\begin{equation}\label{eq:vort_incomp4} 
	\frac{\partial{\xi}_z} {\partial t}+\left(\vec{v}_d \cdot \vec \nabla\right)
	{{\xi}_z}=-{\frac{g}{\rho_{d0}}}{\frac{\partial{\rho_{d1}}} {\partial x}}
	+{\frac{\partial}{\partial x}}\left({\frac{\psi_{y}}{\rho_d}}\right)
	-{\frac{\partial}{\partial y}}\left({\frac{\psi_{x}}{\rho_d}}\right){.}   
	\end{equation}
	We have used the  LCPFCT method (Boris {\it et al.} \cite{boris_book}) to evolve the coupled set of Eqs. (\ref{eq:cont_incomp3}),~  (\ref{eq:psi_incomp3}) and (\ref{eq:vort_incomp3}) for various kinds of density profiles. This method is based on a finite difference scheme associated with the Flux-Corrected Algorithm. The velocity at each time step is updated by using the Poisson's equation ${\nabla^2}{\vec{v}_d}=-{\vec {\nabla}}{\times}{\vec \xi}$, which  has been solved by using the FISPACK 
	package \cite{swarztrauber1999fishpack}. In the subsequent sections we will present the results of 2D numerical simulations of gravitational and buoyancy-driven instabilities and discuss how the growth rate of these instabilities gets suppressed with increasing coupling strength as expected analytically in Eqs.~(\ref{eq:strong_smooth_dispersion})-(\ref{eq:strong_sharp_dispersion}). Boundary conditions are periodic in the horizontal direction (x-axis) and absorbing (non-periodic a) along the vertical (y-axis) direction throughout for all  simulation studies.
	\section{Rayleigh Taylor instability}\label{num_gd_bd}
	
\begin{SCfigure}[][h]
		\includegraphics[width=0.5\textwidth]{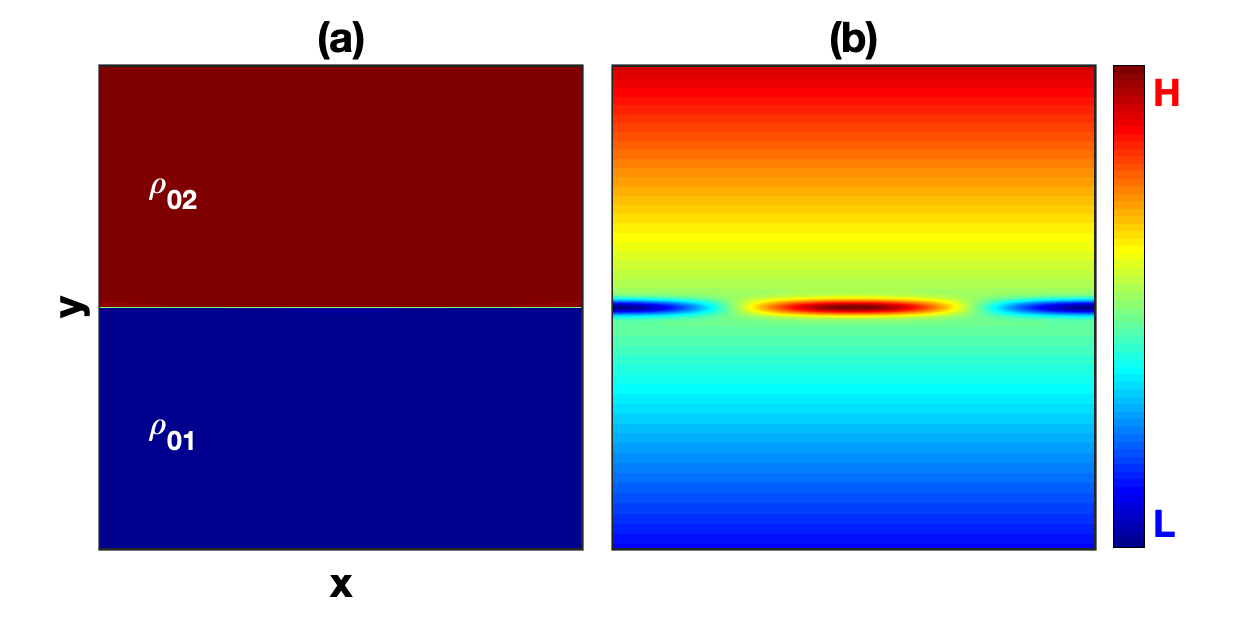}
		\caption{Initial density profiles for Rayleigh-Taylor at time t=0. (a) sharp density interface profile and (b) gradually increasing density gradient along the vertical y-axis; where {\bf H} stands for heavy density and {\bf L} for lighter one.}
		\label{fig:profilert3}	  
	\end{SCfigure}

	We consider two types of inhomogeneous density profiles, namely (A) with sharp interface between two different densities where heavier fluid $\rho_{d02}$ (upper) is supported by lighter one $\rho_{d01}$ (lower) as shown in Fig.~\ref{fig:profilert3}(a) and (B) with gradually increasing density gradient along the y-axis as shown in Fig.~\ref{fig:profilert3}(b). In colorbar, letter {\bf {H}} is the acronym of the heavy density region and {\bf L} stands for lighter density region. 
	\FloatBarrier
	For both cases A and B, the system length is $lx=ly=2{\pi}$ units and density gradient is chosen along the vertical y-axis opposite to gravitational acceleration $g(=10)$. A sufficient electric field is applied  for equilibrium balance against gravity to levitate the dust particles, we had a  discussion about it in  Section~\ref{anal_description}. 
	\subsection{Sharp interface}
	\label{Sub:Sharp interface_num}
	
\begin{figure}[h]
		\includegraphics[width=1.0\textwidth]{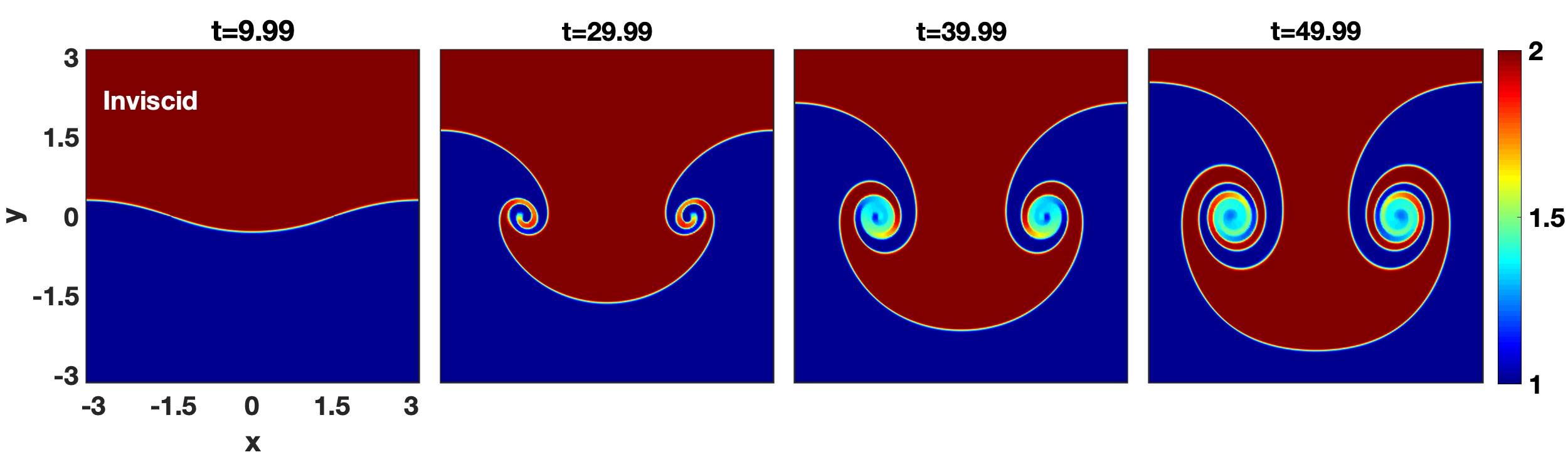}
		\caption{Time evolution of  RT instability for inviscid HD fluid at the sharp interface of two different densities. The heavy fluid (upper half) penetrate into the lighter fluid  (lower half) and the lighter fluid rises above into the domain of heavy fluid. The RT instability perturbations first grows exponentially then saturate to form the bubbles and spikes types of structures.}
		\label{fig:density_sharp_invicid}	         
	\end{figure}

\begin{figure}[h]
		\includegraphics[width=1.0\textwidth]{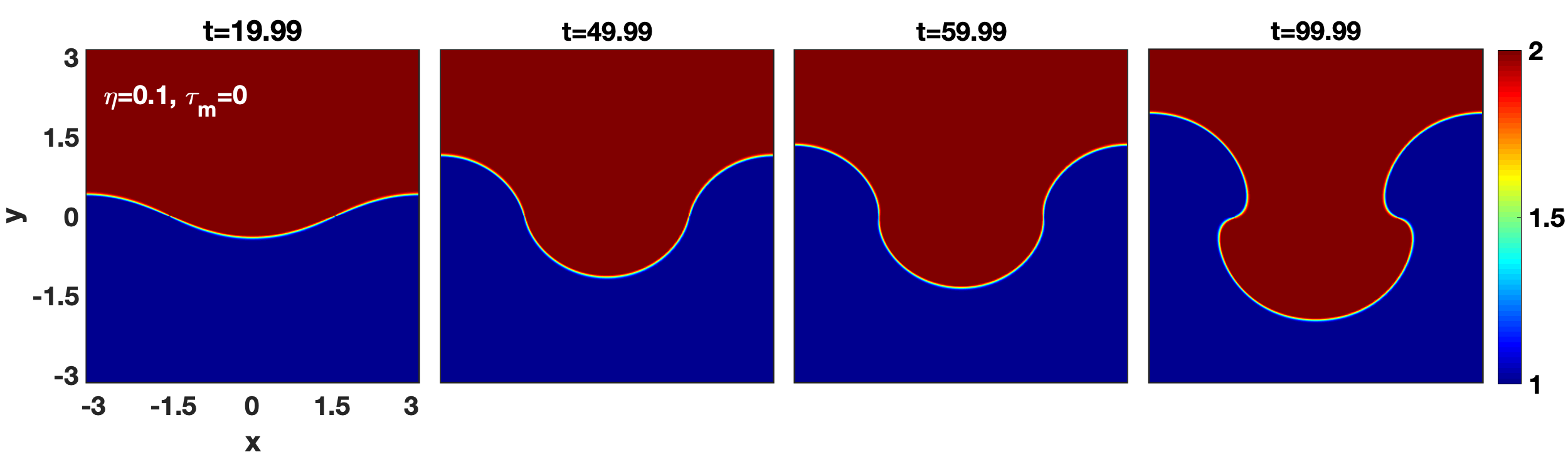}
		\caption{Time evolution of  RT instability for pure viscous HD fluid ($\eta$= 0.1; $\tau_m$=0) at the sharp interface of two different densities. Here, due to the viscous damping force, the RT instability grows at a considerably slower pace compared to the inviscid HD fluid (Fig.~\ref{fig:density_sharp_invicid}).}	
		\label{fig:z_density_sharp_hd_eta0p1}	         
	\end{figure}

\begin{figure}[h]
		\includegraphics[width=1.0\textwidth]{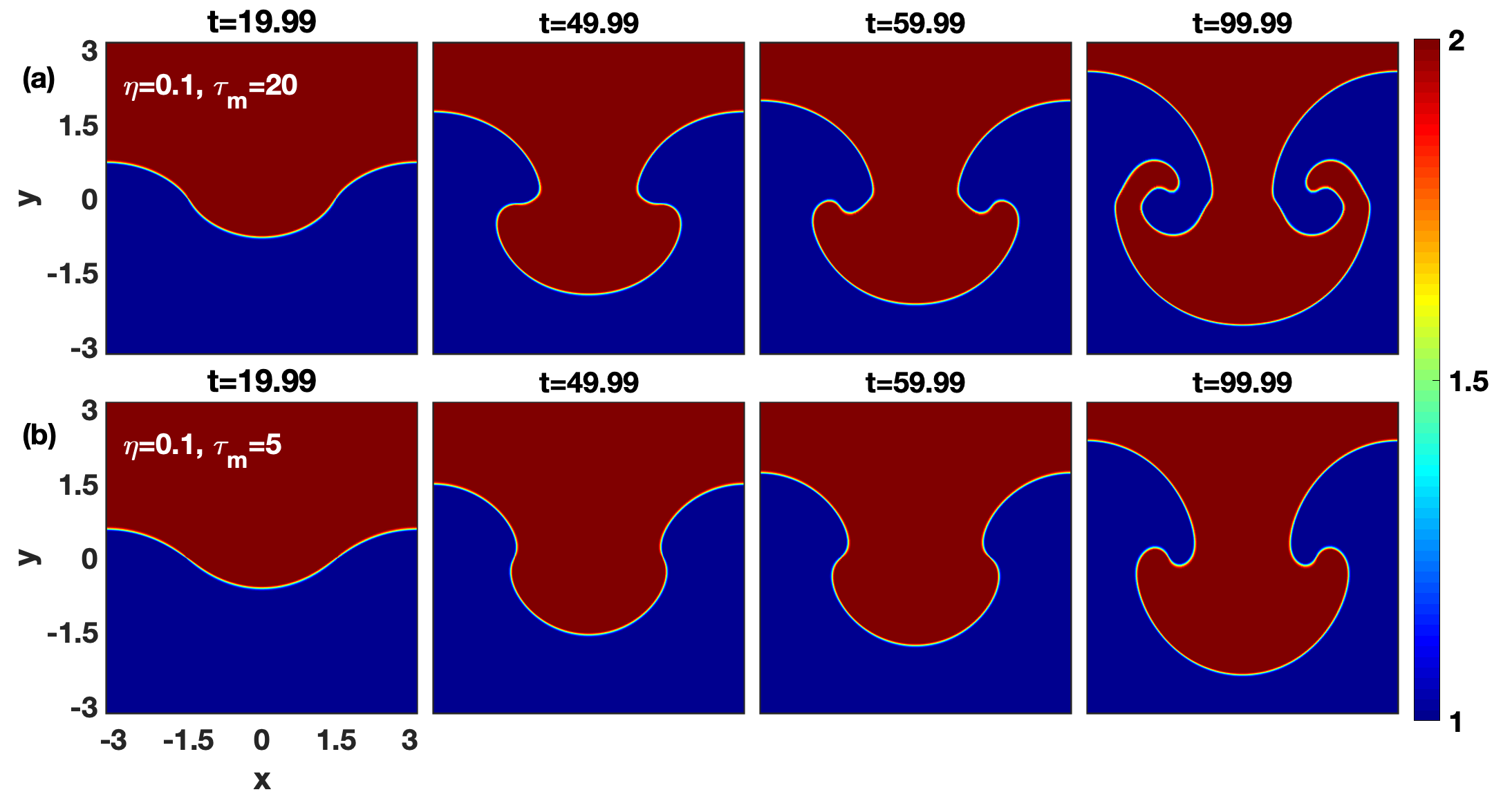}
		\caption{The growth of  RT instability at the sharp interface of two viscoelastic fluids of different densities. (a) $\eta$= 0.1; $\tau_m$=20, and (b) $\eta$= 0.1; $\tau_m$=5. RT instability gets weaker with increasing coupling strength i.e. ${\eta}/{\tau_m}$ as expected analytically from Eq.~(\ref{eq:strong_sharp_dispersion}). Moreover, a comparison between viscous HD fluid (Fig.~\ref{fig:z_density_sharp_hd_eta0p1}) and viscoelastic fluids (Fig.~\ref{fig:density_GHDsharp}) evidents that the appearance of elasticity $\tau_m$ speed up the growth of RT instability by reducing the viscous fraction ($\eta$/$\tau_m$).}
		\label{fig:density_GHDsharp}
	\end{figure}

\begin{SCfigure}[][h]
		\includegraphics[width=0.6\textwidth]{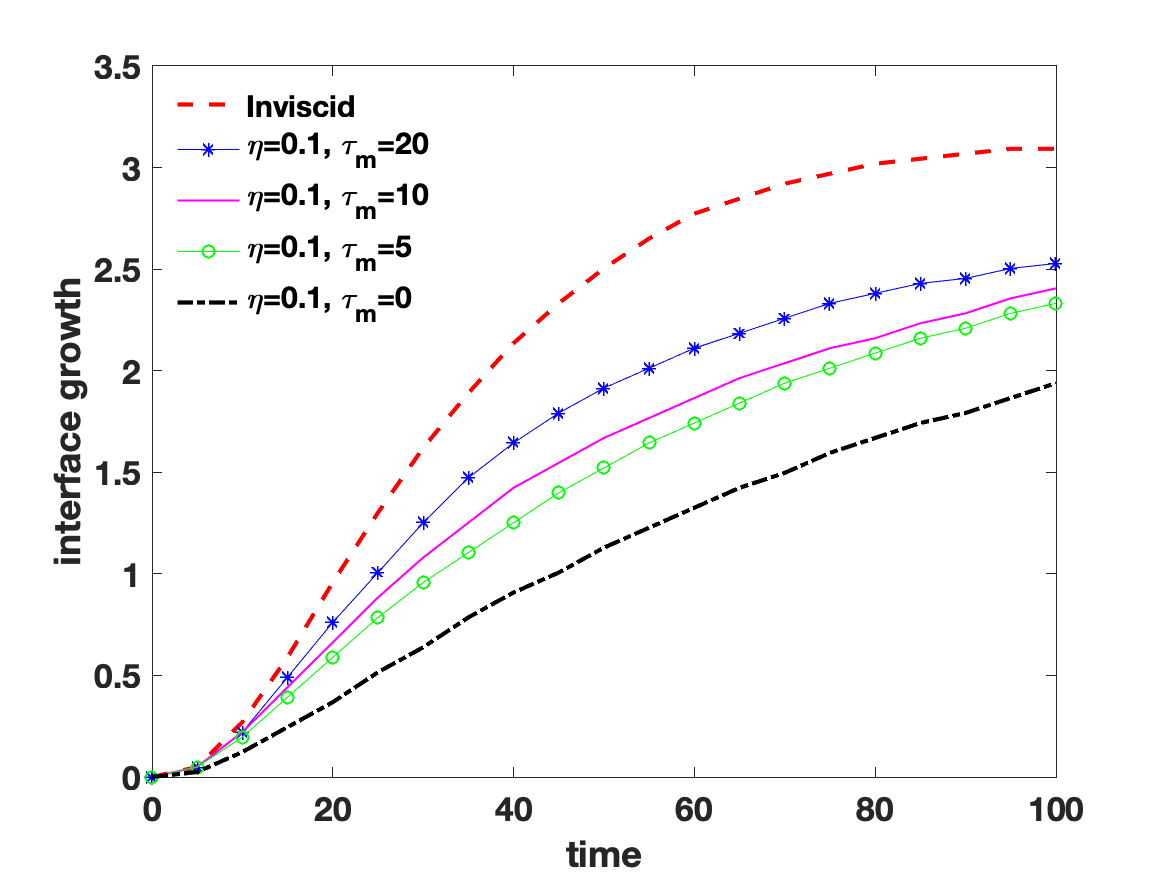}
		\caption{The density interface growth (distance from the initial interface) in vertical downward direction with time. The RT instability gets weaker with increasing the ratio ${\eta}/{\tau_m}$ as expected analytically from Eq.~(\ref{eq:strong_sharp_dispersion}).}
		\label{fig:den_interface_growth}	         
	\end{SCfigure}

	For case A (Fig.~\ref{fig:profilert3}(a)), we consider a system which consists of  two incompressible fluids of constant densities $\rho_{d01}$=1 for $-{\pi}\leq y \leq 0$ (lower half) and $\rho_{d02}$=2 for $0=y \leq {\pi}$ (upper half).   The denser fluid  $\rho_{d02}$ is positioned above the lighter  fluid of density $\rho_{d01}$. This equilibrium configuration remains stable in absence of gravity. To hasten the evolution of our equilibrated system under gravity, we impose a small amplitude  sinusoidal perturbation on the density profile at the interface ($y=0$)  $i.e.$
	\begin{equation}\label{eq:rt_perb}
	\rho_{d1}={\phi_0}{cos({\mathrm{k}}_xx)}exp(-y^2/{\epsilon^2}){.}
	\end{equation}
	Here ${\mathrm{k}}_x$ is the wavenumber associated with the perturbation. Also, the parameter $\epsilon$ defines  the width of the perturbation along $y$. For our simulations we have chosen   these parameters as ${\phi_0}=0.01$, $\epsilon=0.1$ and ${\mathrm{k}}_x=1$. The total density is $\rho_d=\rho_{d0s}+\rho_{d1}$, where the suffix $s$ corresponds to $1$ and $2$ which are labeled for the low and high density of the dust fluid, respectively.  The evolution of RT instability is reproduced for the inviscid HD fluid from our code (see Fig.~\ref{fig:density_sharp_invicid}). 
	\FloatBarrier
	The initial perturbation on the interface was given at ${\mathrm{k}}_x=1$ which is observed. The heavy fluid is observed to penetrate into the lighter fluid and the lighter fluid rises above into the domain of heavy fluid.  Subsequently, the falling heavy fluid develops rolls at the edges acquiring a mushroom like profile. These rolls grow up with time as the heavy  fluid  keeps falling with time.  Such an evolution of RT instability is a well known result of HD fluid \cite{hosseini2017isogeometric, talat2018phase}, thereby substantiating our numerical simulation. In order to see the pure viscous damping effect on the growth of RT instability, we have  simulated the viscous HD fluid of $\eta= 0.1$ with the value of $\tau_m = 0$  shown in Fig.~\ref{fig:z_density_sharp_hd_eta0p1}.
	\FloatBarrier
	Here, the instability grows at a considerably slower pace compared to the inviscid HD fluid (Fig.~\ref{fig:density_sharp_invicid}) because of the viscous damping force. In the interest of elastic effects on the growth of instability we have also considered  two viscoelastic  fluids  with two different  values of $\tau_m $ (20 and 5) for  the same value of $\eta = 0.1$ (viscous damping is similar) shown in Fig.~\ref{fig:density_GHDsharp}.  
	A comparison between  viscous HD fluid (Fig.~\ref{fig:z_density_sharp_hd_eta0p1}), viscoelastic fluids (Fig.~\ref{fig:density_GHDsharp}) evidents that the RT instability in the dust fluid phenomenologically similar to the HD case as observed in dusty plasma experiment by Pacha {\it et al.} \cite{pacha2012observation}. It can also be noticed that the appearance of elasticity $\tau_m$ speed up the growth of RT instability  by reducing the viscous fraction ($\eta$/$\tau_m$). Such reducing effect of elasticity on RT instability has also been observed in many studies~\cite{biot1965mechanics,biot1965theory, ode1966gravitational, poliakov1993explicit,naimark1994gravitational,kaus2007effects,boffetta_mazzino_musacchio_vozella_2010}.  Or the RT instability gets weaker with increasing the value of coupling strength ($\eta$/$\tau_m$) as expected analytically from Eq.~(\ref{eq:strong_sharp_dispersion}). In Fig.~\ref{fig:den_interface_growth}, for the quantitative comparison,  the density interface growth for all these has also been evaluated and found to similar with the above observations.
	\FloatBarrier
	\subsection{Gradual variation of density profile}
\begin{figure}[h]
		\includegraphics[width=1.0\textwidth]{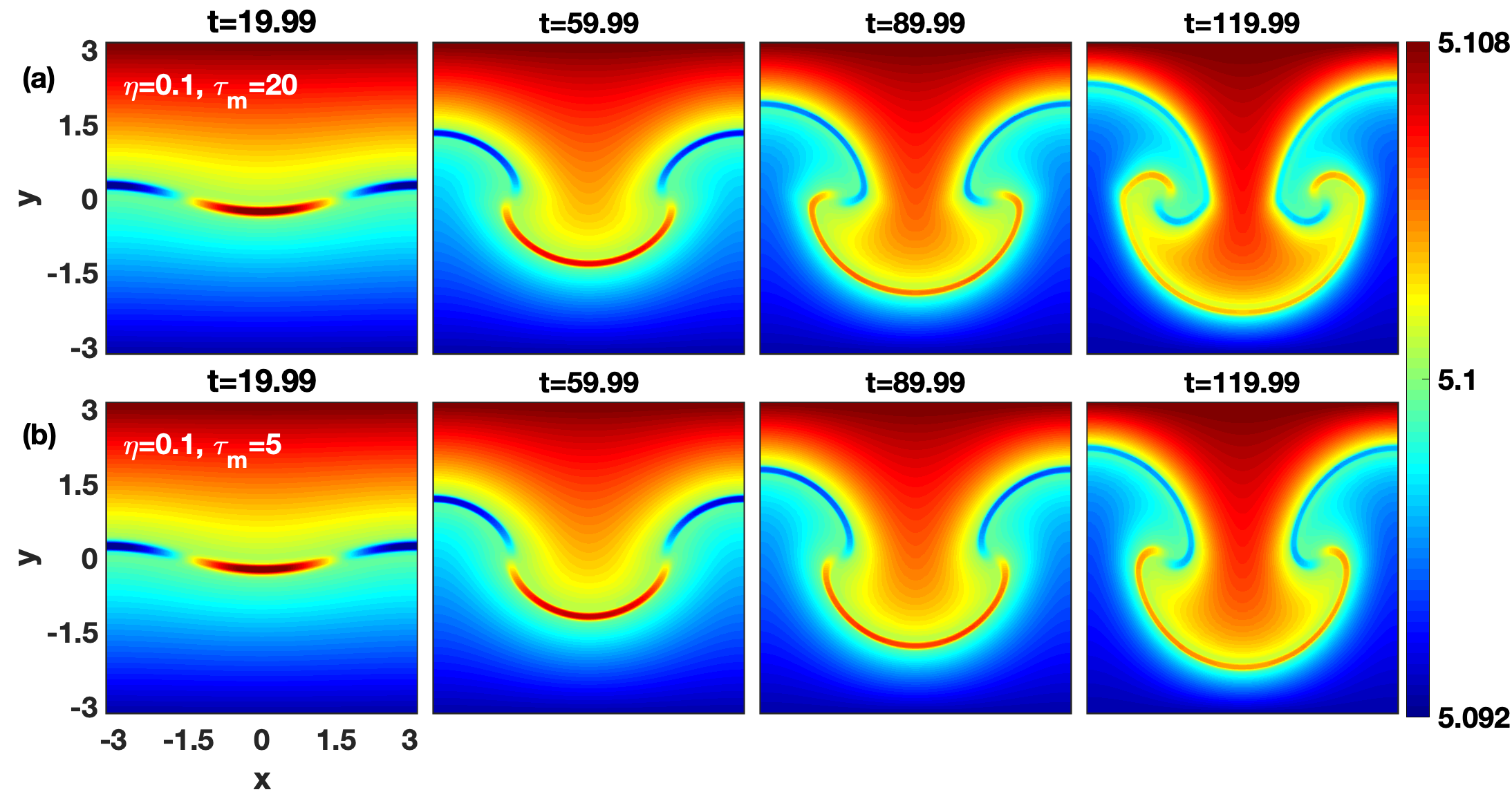}
		\caption{The time evolution of  RT instability for two viscoelastic fluids having gradually increasing densities against the gravity. (a) Viscoelastic fluid with $\eta$= 0.1; $\tau_m$=20, and (b) viscoelastic fluid with $\eta$= 0.1; $\tau_m$=5. RT instability gets weaker with increasing coupling strength i.e. ${\eta}/{\tau_m}$ as analytically suggested by Eq. (\ref{eq:strong_smooth_dispersion}).}
		\label{fig:density_smooth_eta0p1_tm520}	                
	\end{figure}

	Next, the density of the dust fluid is stratified against gravity and has a gradually changing profile as shown in  Fig.~\ref{fig:profilert3}(b).  
	\begin{equation} \label{eq:smooth_rt_prof}
	\rho_{d}={\rho_{d0}}+{\phi_0}exp({\sigma}y)+{\rho_{d1}}{.}
	\end{equation}
	Here $\rho_{d0}$ is constant background density. The value of parameters taken for the present case are $\rho_{d0}$=5, amplitude of inhomogeneity $\phi_0$=0.1 and $\sigma$=0.025 decide the ramp of inhomogeneity in background density. The sinusoidal perturbation ${\rho_{d1}}$ (see Eq.~\ref{eq:rt_perb}) is imposed  at y = 0, which can also be clearly seen in Fig.~\ref{fig:profilert3}(b).  In Fig.~\ref{fig:density_smooth_eta0p1_tm520}, we have compared two different viscoelastic fluids with same $\eta$  ({${\eta}=0.1, {\tau_m}=20, {\eta}/{\tau_m}$=0.005} for Fig.~\ref{fig:density_smooth_eta0p1_tm520}(a), ({${\eta}=0.1, {\tau_m}=5, {\eta}/{\tau_m}$=0.02} for Fig.~\ref{fig:density_smooth_eta0p1_tm520}(b)).  From the comparison of Fig.~\ref{fig:density_smooth_eta0p1_tm520}(a) and Fig.~\ref{fig:density_smooth_eta0p1_tm520}(b), it is clear that at each time step, the growth of RT instability gets weaker with increasing ${\eta}/{\tau_m}$ as suggested by Eq. (\ref{eq:strong_smooth_dispersion}) or the reducing effect of elasticity to the viscous fraction ($\eta$/$\tau_m$) enhances the RT growth. The elastic behaviour of the system is known to produce TS waves in the medium. Similar observations have been made for the sharp  density gradient profiles (Subsection \ref{Sub:Sharp interface_num}). It is clear that in all these cases the evolution is identical suggesting that the shear wave associated with strong coupling is operational.
	\FloatBarrier
	\section{Buoyancy-driven evolution}
	
\begin{SCfigure}[][h]
		\includegraphics[width=0.6\textwidth]{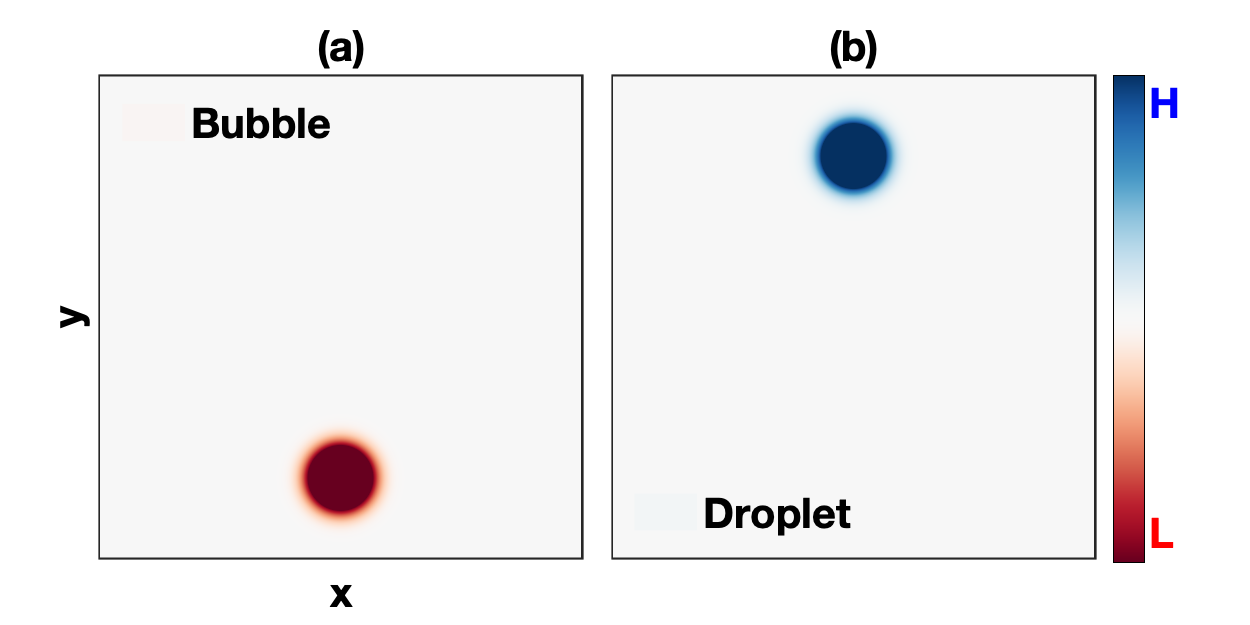}
		\caption{Initial densities' profiles  at time t=0. (a) Bubble  of less density inside the heavier fluid and (b) droplet of high density inside the lighter fluid.  In colorbar, letter {\bf {H}} is the acronym of the heavy density region and {\bf L} stands for lighter density region.} 
		\label{fig:initialprofb-d}	        
	\end{SCfigure}

\begin{SCfigure}[][h]
		\includegraphics[width=0.6\textwidth]{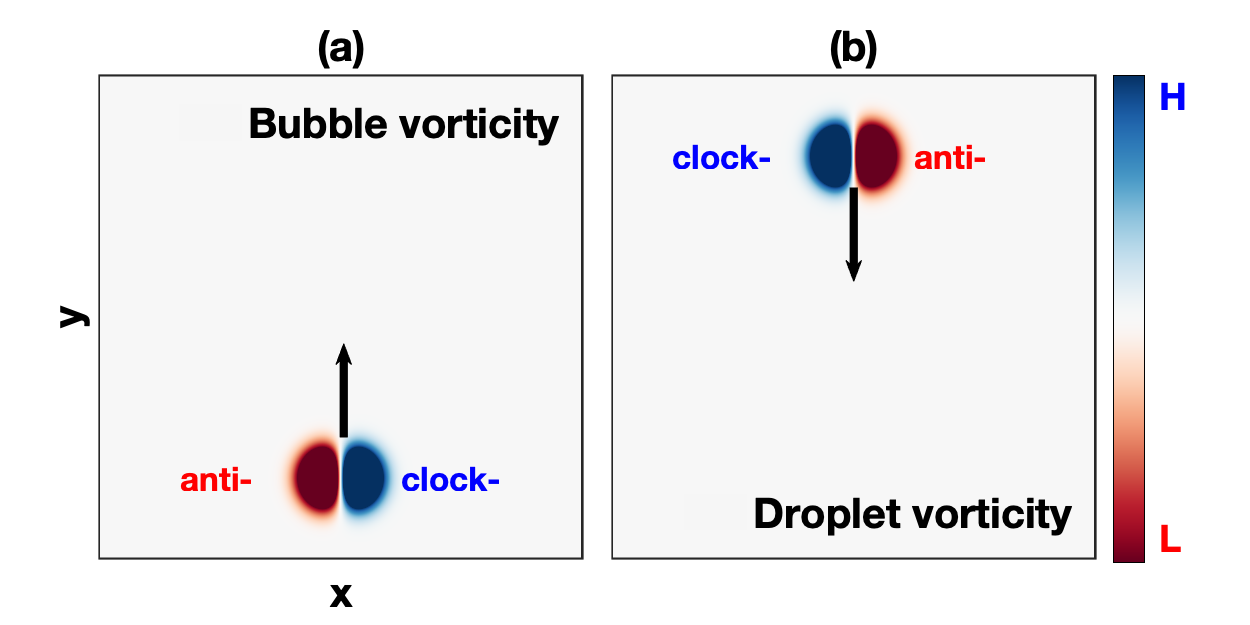}
		\caption{Schematic vorticity diagram (a pair of counter counter-rotating lobes) of the bubble  and droplet. (a) Dipolar vorticity  due to the buoyant force on bubble w.r.t. Fig.~\ref{fig:initialprofb-d}(a). Here, the net rotation causes  the motion in vertical upward direction, indicated by vertical upward arrow. Similarly, vorticity in (b) w.r.t. Fig.~\ref{fig:initialprofb-d}(b) causes the net downward flow for the droplet.} 
		\label{fig:rising_bbl_cause}	      
	\end{SCfigure}

	We consider the two types of density inhomogeneity profiles, namely (A) an initially static and circular bubble whose density is less than that of the surrounding quiescent Newtonian/viscoelastic fluid as shown in  Fig.~\ref{fig:initialprofb-d}(a) and (B) initially static and a circular droplet whose density is higher than that of the surrounding medium as shown in Fig.~\ref{fig:initialprofb-d}(b). In both the cases the variation in density is symmetric about an axis which is perpendicular to simulation plane.
	\FloatBarrier
	For both cases, we have considered a system of length $lx=ly=12{\pi}$ units, approximating the gravitational acceleration as $g=10$. Under the influence of gravity, the less dense regions (bubble)  have a tendency to rise  upwards, whereas the higher density (droplet)  sinks against the background low density fluid.  Now let us discuss the equations of model to develop the some simple qualitative understanding of simulations results. For an inviscid flow, the vorticity Eq.~(\ref{eq:vort_incomp3})/Eq.~(\ref{eq:vort_incomp4}) in hydrodynamic limit, becomes 
	\begin{equation}\label{eq:fluid1} 
	\frac{\partial{\xi_{z}}} {\partial t}+\left(\vec{v}_d \cdot \vec \nabla\right)
	{\xi_{z}}={\frac{1}{\rho_{d0}}}{{\nabla}{\times}{\rho_{d1}}{\vec{g}}}=-{\frac{g}{\rho_{d0}}}{\frac{\partial{\rho_{d1}}} {\partial x}}{.} 
	\end{equation}
	The spatial variation in density $i.e.$ ~$-{({g}/{\rho_{d0}})}{({\partial{\rho_{d1}}}/{\partial x})}$ acts as a buoyant force. In order to simulate the case of rising/falling bubble/droplet, we have considered a Gaussian density profile given by,
	\begin{equation}\label{eq:rising_bbl} 
	\rho_{d1}={\rho^{\prime}_{0}}exp\left(\frac{-\left({\left(x-x_{c}
			\right)^2+(y-y_{c})^2}\right)}{a^2_{c}}\right){,}
	\end{equation}
	where $a_c$ is the bubble/droplet core radius. As the simulation begins, the buoyant force comes into a picture which in turn induces the dipolar vorticity (a pair of counter rotating vortices)  for our considered Gaussian density profile $i.e.$
	\begin{equation}\label{eq:vort_rspkt_dens} 
	-{\frac{\partial{\rho_{d1}}} {\partial x}}=2{\left(x-x_{c}\right)}{ \rho_{d1}}{.}
	\end{equation}
	The net resultant of counter rotating  pair of vortices will decide the net flow direction of the medium~\cite{horstmann2014wake} $i.e$ vertical upward (bubble) or vertical downward (droplet). Figure~\ref{fig:rising_bbl_cause}, corresponds to the Fig.~\ref{fig:initialprofb-d}, shows the schematic diagram of vorticities (using Eq.~(\ref{eq:vort_rspkt_dens})), it is clear that two similar counter-rotating lobes cause the net propagation of bubble/droplet  in the vertical upward/downward direction (indicated by arrows).
	\FloatBarrier
	While in viscoelastic fluids, in addition to the buoyant force term, the vorticity Eq.~(\ref{eq:vort_incomp3})/Eq.~(\ref{eq:vort_incomp4})  includes  ${\nabla}{\times}{({\vec \psi}/{\rho_d})}$ as second term (RHS) which incorporate the shear waves into the medium moving with phase velocity $\sqrt{{\eta}/{{\rho_d}{\tau_m}}}$.
	\subsection{Rising bubble }
\begin{figure}[h]
		\includegraphics[width=1.0\textwidth]{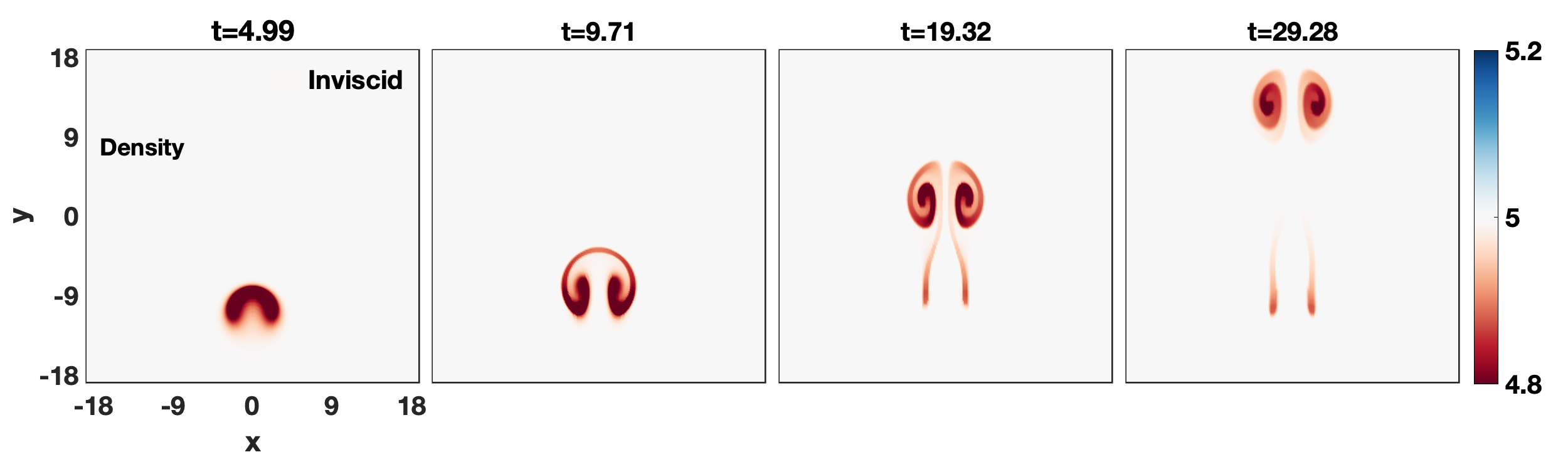}
		\caption{Time evolution of bubble density (see Fig.~\ref{fig:bbl_vort_g10_all_b5_fluid} for the respective vorticity) profile for inviscid HD fluid.}
		\label{fig:bbl_den_g10_fluid_b5}
	\end{figure}

\begin{figure}[h]
		\includegraphics[width=1.0\textwidth]{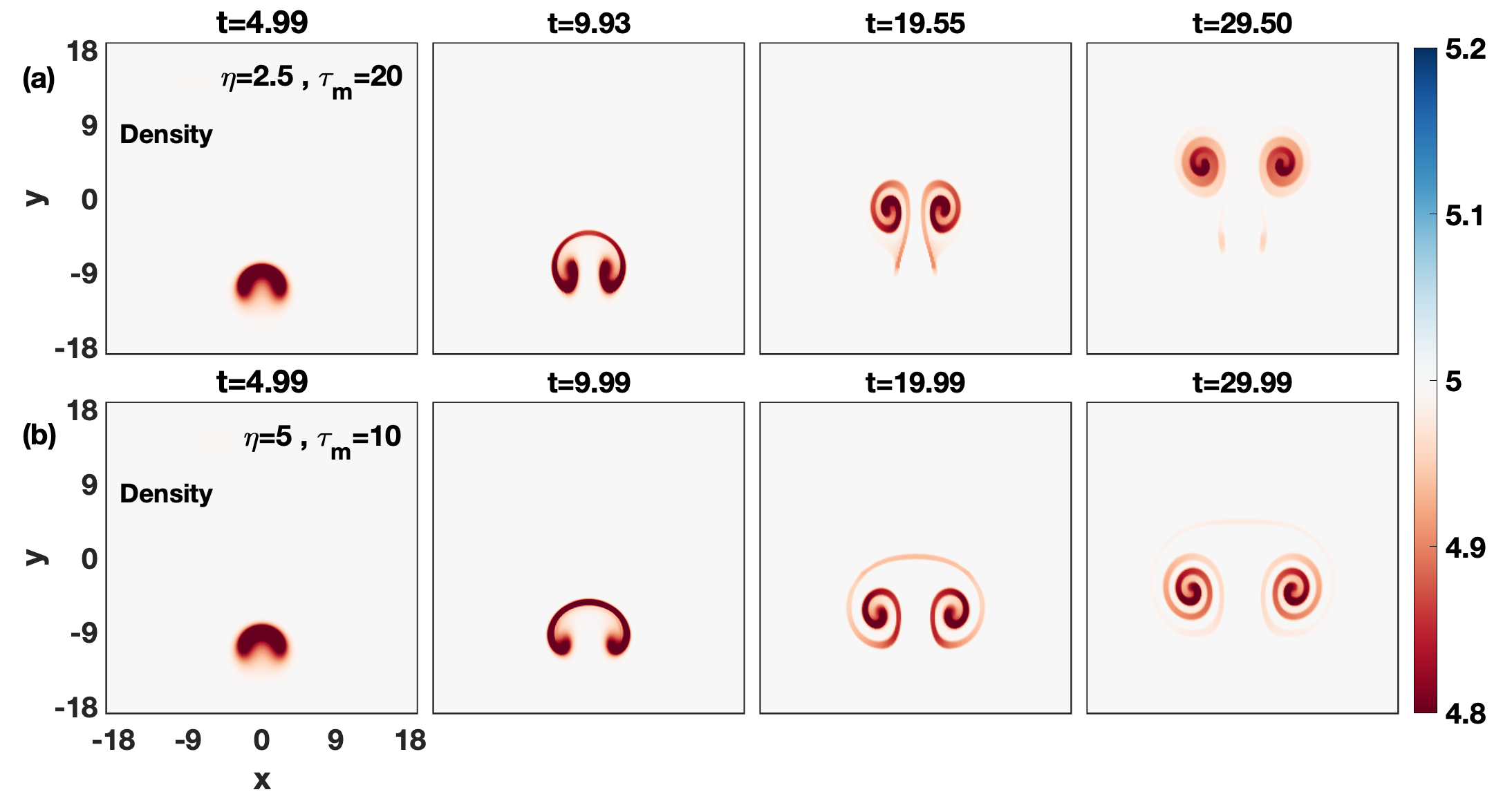}
		\caption{Evolution of bubble density profile (see Fig.~\ref{fig:bbl_vort_g10_all_b5} for the respective vorticity evolution) in time for two viscoelastic fluids (a) $\eta$= 2.5; $\tau_m$=20, and (b) $\eta$= 5; $\tau_m$=10. The growth of bubble gets reduced (rising rate) with increasing coupling strength.}
		\label{fig:bbl_den_g10_510_b4}	                
	\end{figure}

\begin{figure}[h]
		\includegraphics[width=1.0\textwidth]{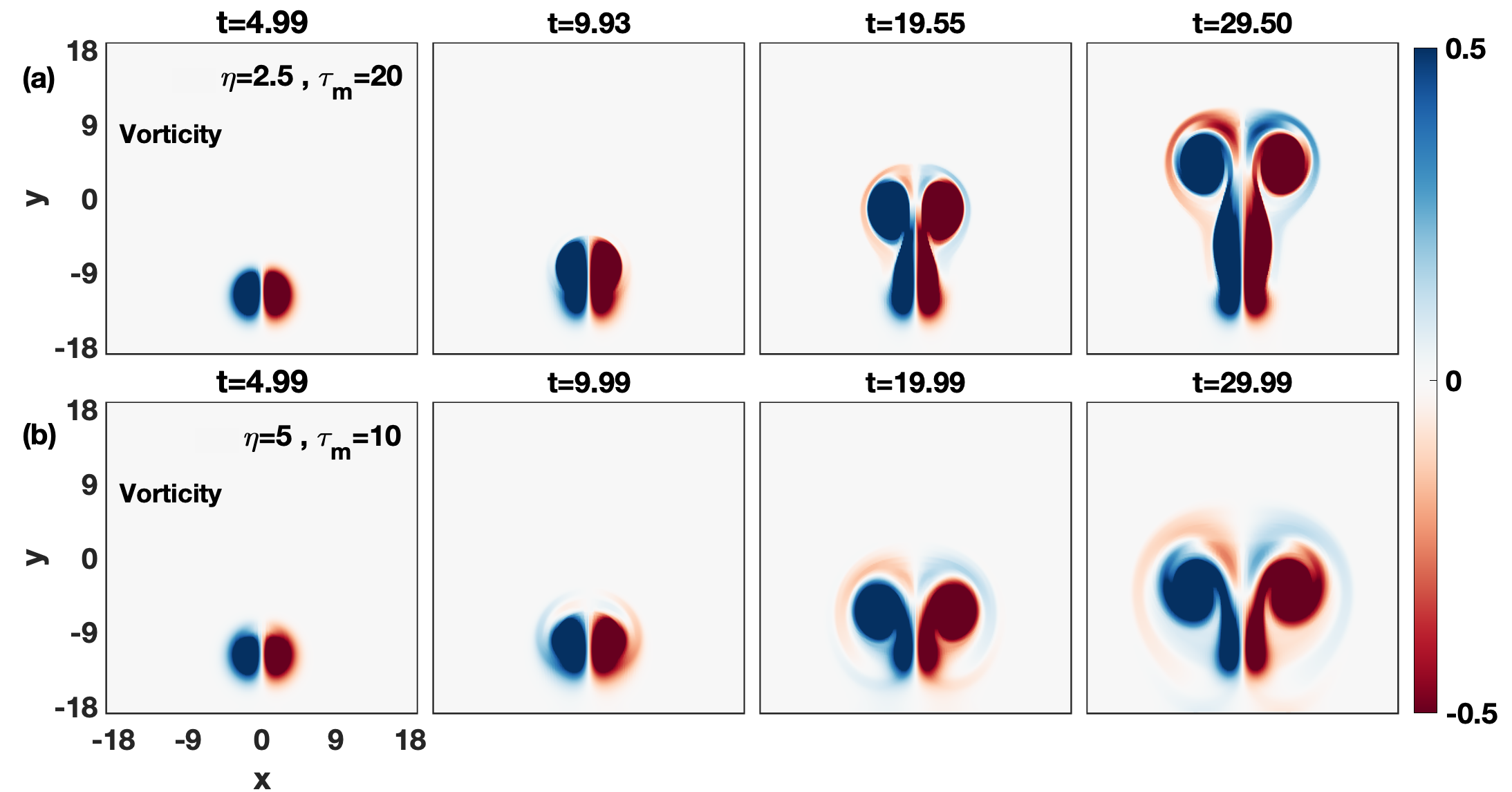}
		\caption{Time evolution of bubble vorticity profile (see Fig.~\ref{fig:bbl_den_g10_510_b4} for the respective density evolution) for two viscoelastic fluids (a) $\eta$= 2.5; $\tau_m$=20, and (b) $\eta$= 5; $\tau_m$=10.  There is emission of TS wave surrounding the vorticity lobes whose phase velocity is proportional to the coupling strength. It is clear that the  increase of phase velocity  of  TS  waves with coupling strength suppresses the BD instability ($i.e$ reduction in rising rate of lobes).}
		\label{fig:bbl_vort_g10_all_b5}	 
	\end{figure}

\begin{figure}[h]
		\includegraphics[width=1.0\textwidth]{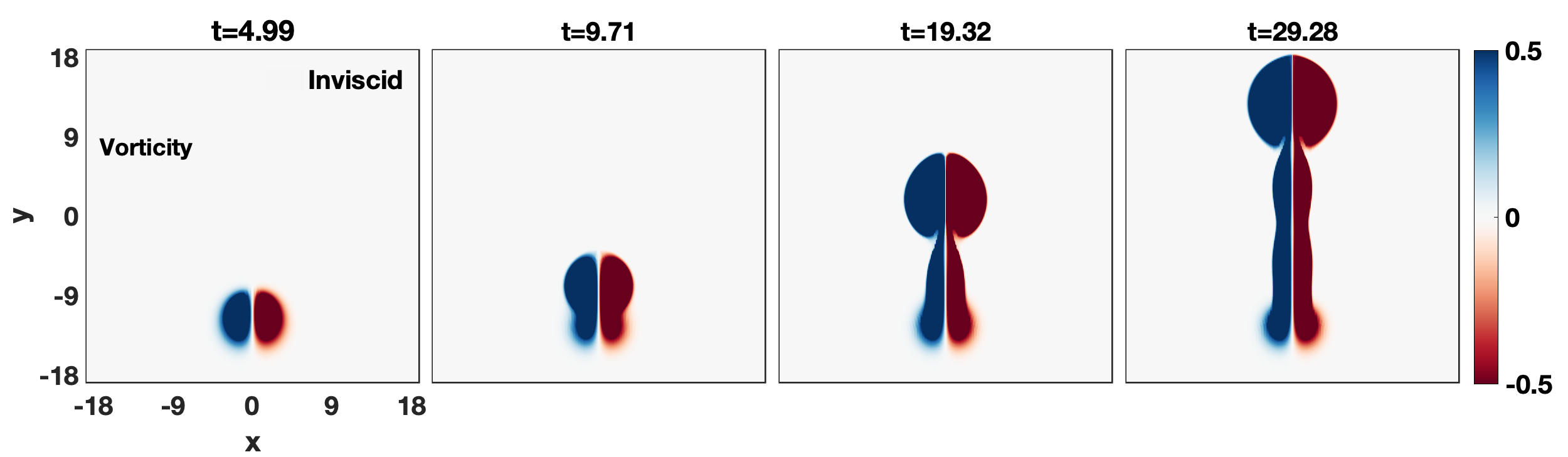}\caption{Evolution of bubble vorticity profile in time for inviscid HD fluid (see Fig.~\ref{fig:bbl_den_g10_fluid_b5} for the respective density).}
		\label{fig:bbl_vort_g10_all_b5_fluid}	
	\end{figure}

\begin{SCfigure}[][h]
		\includegraphics[width=0.5\textwidth]{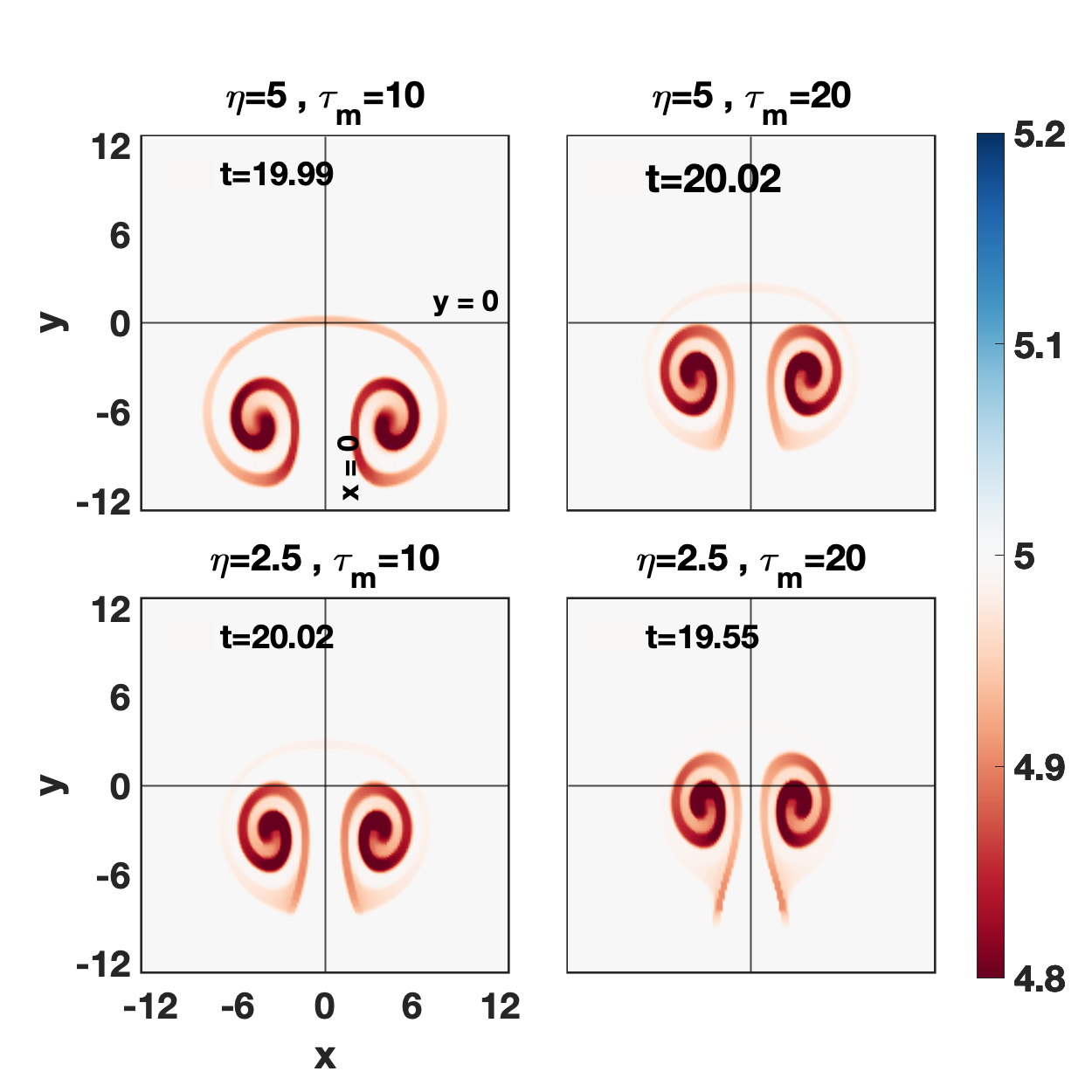}
		\caption{Viscosity $\eta$ vs elasticity $\tau_m$. The snapshots of bubble density  profiles at a particular timestep (see Fig.~\ref{fig:z_vis_elsticity_vort2b2} for the respective vorticity). The BD instability gets weaker with increasing viscosity (in the vertical direction with a fixed value of $\tau_m$), and elasticity only reduces the viscous effects (along the horizontal direction with a fixed value of $\eta$).}
		\label{fig:z_vis_elsticity_dens2b2}	       
	\end{SCfigure}

\begin{SCfigure}[][h]
		\includegraphics[width=0.5\textwidth]{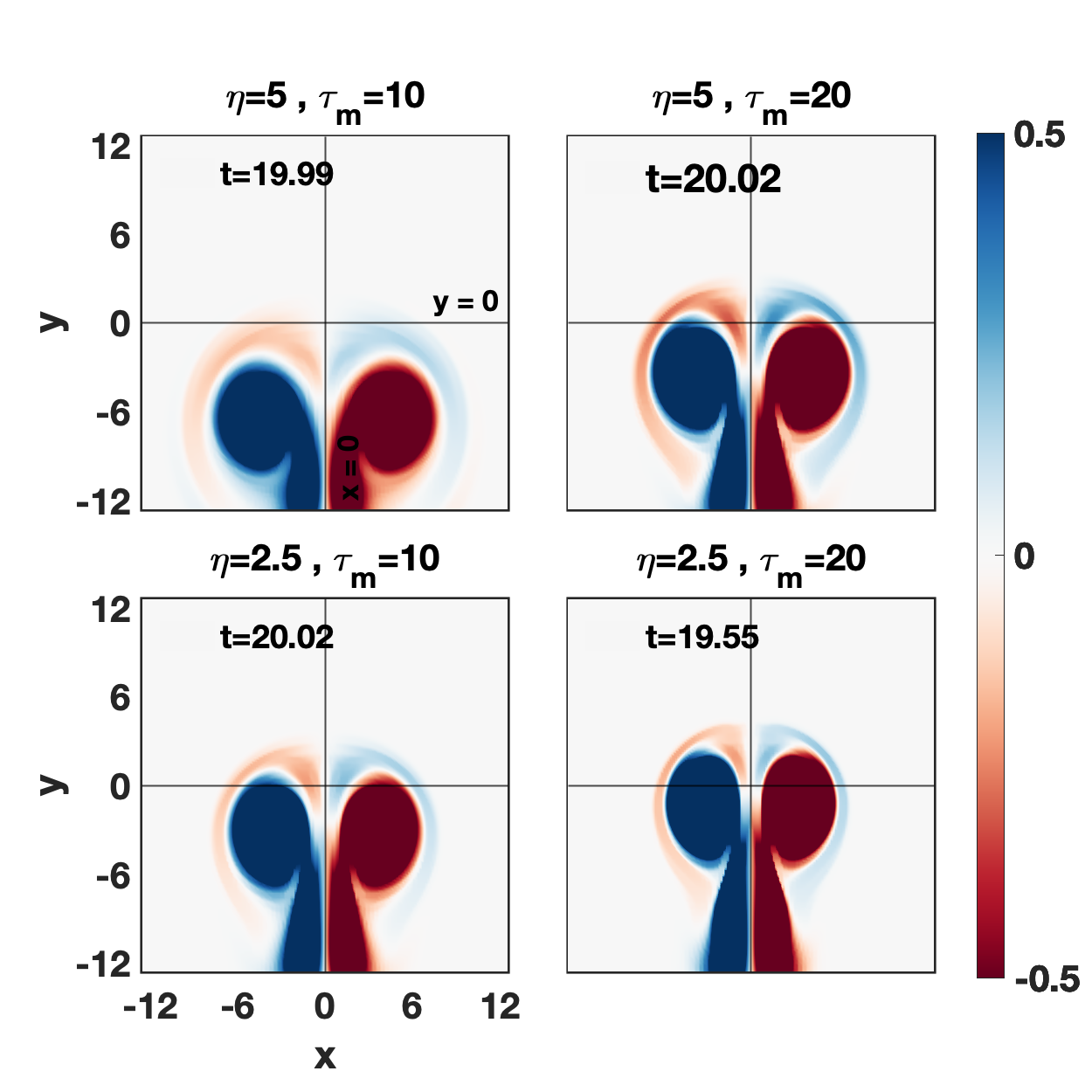}
		\caption{Viscosity $\eta$ vs elasticity $\tau_m$. The snapshots of bubble vorticity profiles at a particular timestep (see Fig.~\ref{fig:z_vis_elsticity_dens2b2} for the respective density). The BD instability gets weaker with increasing viscosity (in the vertical direction with a fixed value of $\tau_m$), and elasticity only reduces the viscous effects (along the horizontal direction with a fixed value of $\eta$).}
		\label{fig:z_vis_elsticity_vort2b2}	 
	\end{SCfigure}

	In order to simulate the case of rising bubble the numerical simulation has been carried out for ${a_c}$=2.0, ${\rho^{\prime}_{0}}=-0.5$ and ${x_{c}}=0$, ${y_{c}}=-4\pi$ in above Eq.~\ref{eq:rising_bbl}. The homogeneous background density with ${\rho_{d0}}=5$ has been taken. Thus, the total density is $\rho_d=\rho_{d0}+\rho_{d1}$. Figure~\ref{fig:bbl_den_g10_fluid_b5} displays the time evolution of the bubble density profile of inviscid fluid. Initially at time $t=0$, the bubble is axisymmetric (see Fig.~\ref{fig:initialprofb-d}(a)) and as the system evolves, the bubble  simply rises without any significant change in shape for a short period of  time. At a later stage, the initially circular profile assumes a crescent  shape as evident from the first subplot of Fig.~\ref{fig:bbl_den_g10_fluid_b5}. Further, as time progresses, a mushroom-like structure which is characteristic of RT instability begins to appear along with rolling and mixing at the edge of the bubble. This mushroom-like structure then breaks into two distinct elliptical lobes as evident from the figure. These lobes propagate forward as a single entity leaving behind wake-like structure in background fluid. Next, with the understanding of bubble evolution for HD fluids, the time evolution of bubble for viscoelastic fluids has been performed. 
	\FloatBarrier
	To envisage the effect of coupling strength, we compare one viscoelastic fluid (Fig.~\ref{fig:bbl_den_g10_510_b4}(a)) having ${\eta}=2.5, {\tau_m}=20$ i.e.  $\eta/\tau_m$=0.125 with another viscoelastic fluid  (Fig.~\ref{fig:bbl_den_g10_510_b4}(b)) having higher coupling strength i.e. ${\eta}$=5, ${\tau_m}$=10, $\eta/\tau_m$=0.5.
	\FloatBarrier
	We observe a decrease in rising rate of bubble with increasing coupling strength. Also, the elliptical lobes start moving apart in horizontal direction earlier in stronger coupled medium. At the same time, the rolling rate of lobes increases causing the expansion of lobes and disappearance of the dragging tail. Thus, we observe that similar to the RT instability the growth of bubble gets reduced (rising rate) with increasing coupling strength and this reducing effect can be easily visualised through the respective vorticity evolution given in Fig.~\ref{fig:bbl_vort_g10_all_b5}.
	\FloatBarrier
	In Fig.~\ref{fig:bbl_vort_g10_all_b5}, the vorticity subplots in first and second row correspond to the density profiles shown in Fig.~\ref{fig:bbl_den_g10_510_b4}(a) and Fig.~\ref{fig:bbl_den_g10_510_b4}(b), respectively.  It is evident from Fig.~\ref{fig:bbl_vort_g10_all_b5} that there is emission of TS waves   surrounding the vorticity lobes for visco-elastic fluids and no such TS waves   exist for inviscid HD fluid (Fig.~\ref{fig:bbl_vort_g10_all_b5_fluid}). The emission of waves from either lobe pushes the other lobe in the direction perpendicular to the direction of propagation of the entire  structure and as a result the lobes get well separated with time and proportional to the coupling strength. Besides this lobe separation, the emission of TS wave significantly reduced the strength of dipole thereby impeding the dipole propagation and this reduction in  dipole propagation is also proportional to the coupling strength. The relative observations of Fig.~\ref{fig:bbl_vort_g10_all_b5}(a) and Fig.~\ref{fig:bbl_vort_g10_all_b5}(b) clearly reflect it through the TS enclosure which is larger for the $\eta$= 5; $\tau_m$=10 than  $\eta$= 2.5; $\tau_m$=20. 
	\FloatBarrier
	Thus, we conclude  that the  increase  of  phase  velocity  of  TS  waves with coupling strength suppresses the BD instability ($i.e$ reduction in vertical propagation direction and enhancement in horizontal separation between lobes). To understand the separate effects of viscosity and elasticity on BD growth the snapshots of density/vorticity snapshots at the same timestep $({\approx}20)$ are shown in Fig.~\ref{fig:z_vis_elsticity_dens2b2}/Fig.~\ref{fig:z_vis_elsticity_vort2b2}. From Fig.~\ref{fig:z_vis_elsticity_dens2b2}, the damping effect of viscosity on the BD instability can be easily noticed, and elasticity reduces the viscous effects by reducing the viscous fraction. 
	\FloatBarrier
	The role of TS waves can be visualised from the vorticity snapshots given in Fig. \ref{fig:z_vis_elsticity_vort2b2}.
	\FloatBarrier
	From viscoelastic theory it well known that the strong coupling increases the material properties like its rigidity (giving rise to shear waves). So this extra rigidity should give rise to an extra torque thereby increasing the rotation. This would imply that in strong coupling limit rotation should be faster. This effect can be observed from Fig.~\ref{fig:z_vis_elsticity_dens2b2}/Fig.~\ref{fig:z_vis_elsticity_vort2b2}, in strong coupling limit the expansion of lobes is higher, and disappearance of the dragging tail in density evolution.  Also,  the vortex evolution is similar for the same value of coupling strength (${\eta}/{\tau_m}$). 
	\subsection{Falling droplet}
	
\begin{figure}[h]
		\includegraphics[width=1.0\textwidth]{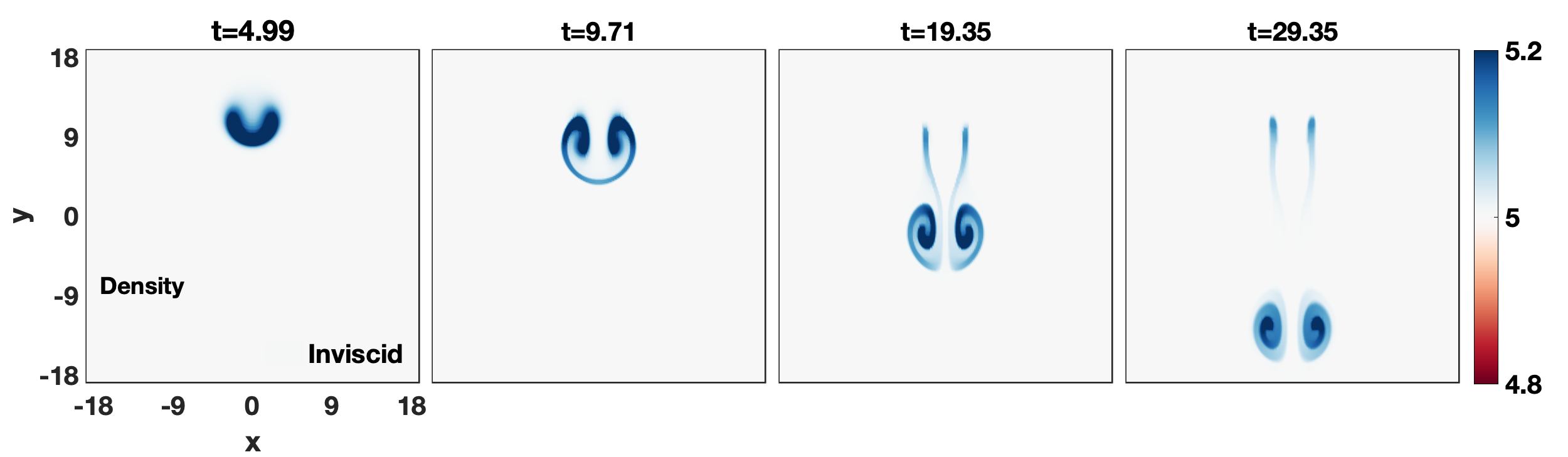} 
		\caption{Time evolution of droplet density (see Fig.~\ref{fig:drop_vort_g10_all_b5_fluid} for the respective vorticity evolution) profile for inviscidHD fluid.}
		\label{fig:drop_den_g10_fluid_b5}	                    
	\end{figure}

\begin{figure}[h]
		\includegraphics[width=1.0\textwidth]{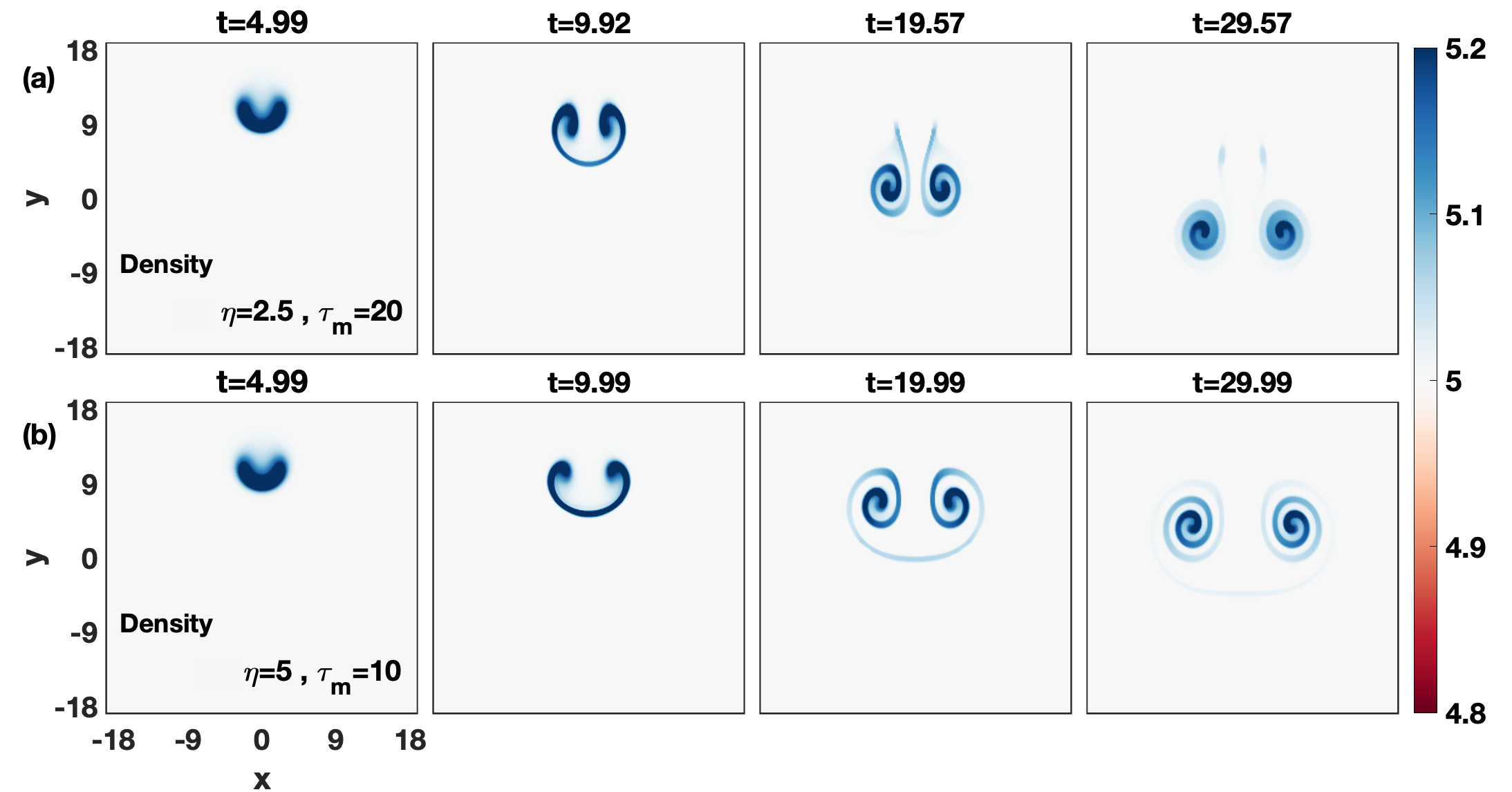}
		\caption{Evolution of droplet density profile (see Fig.~\ref{fig:drop_vort_g10_all_b5} for the respective vorticity evolution) in time for two viscoelastic fluids (a) $\eta$= 2.5; $\tau_m$=20, and (b) $\eta$= 5; $\tau_m$=10. The growth of droplet gets reduced (falling rate) with increasing coupling strength.}
		\label{fig:drop_den_g10_ghd_b5_0p35}	
	\end{figure}

\begin{figure}[h]
		\includegraphics[width=1.0\textwidth]{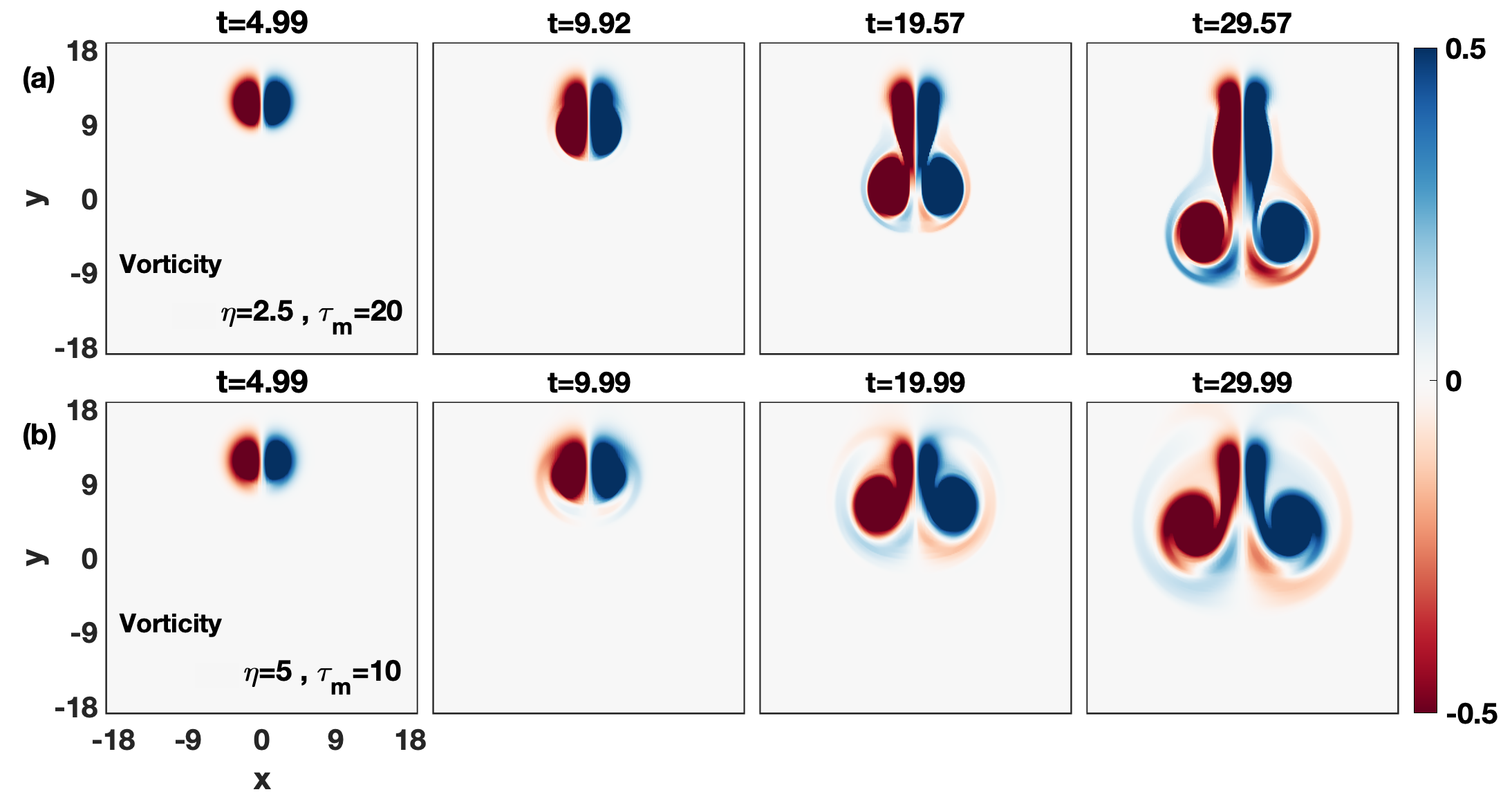} 
		\caption{Time evolution of droplet vorticity profile (see Fig.~\ref{fig:drop_den_g10_ghd_b5_0p35} for the respective density evolution) for two viscoelastic fluids (a) $\eta$= 2.5; $\tau_m$=20, and (b) $\eta$= 5; $\tau_m$=10.  There is emission of TS wave surrounding the vorticity lobes whose phase velocity is proportional to the coupling strength. It is clear that the  increase of phase velocity  of  TS  waves with coupling strength suppresses the BD instability ($i.e$ reduction in falling rate of lobes).}
		\label{fig:drop_vort_g10_all_b5}	  
	\end{figure}

\begin{figure}[h]
		\includegraphics[width=1.0\textwidth]{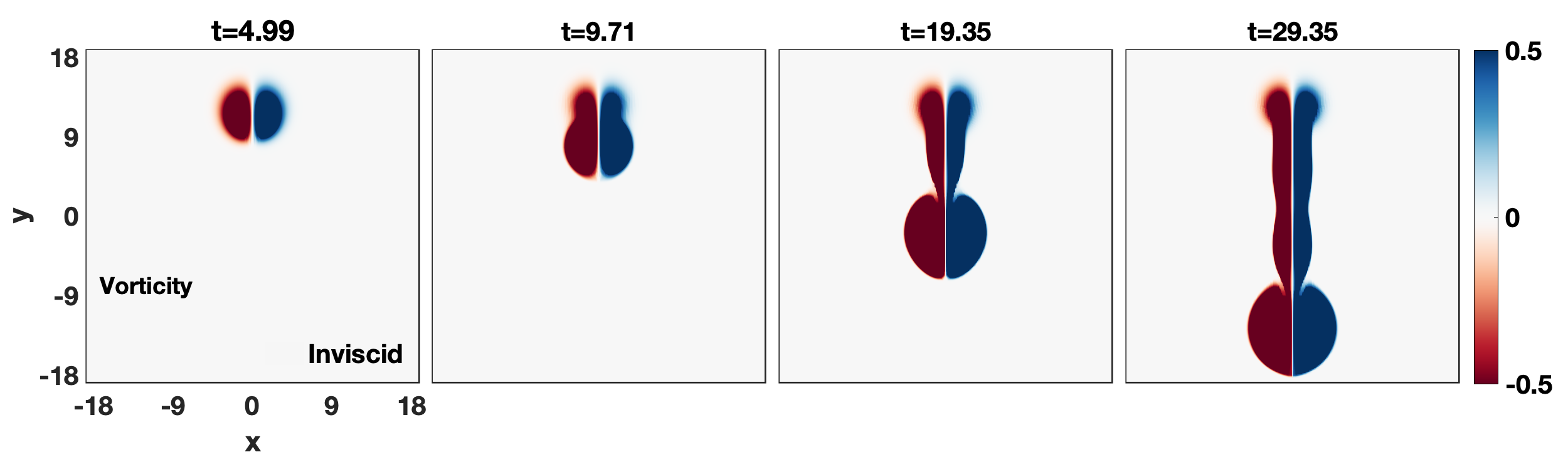} 
		\caption{Time evolution of droplet vorticity (see Fig.~\ref{fig:drop_den_g10_fluid_b5} for the respective density evolution) profile for inviscid HD fluid.}
		\label{fig:drop_vort_g10_all_b5_fluid}	        
	\end{figure}

\begin{SCfigure}[][h]
		\includegraphics[width=0.5\textwidth]{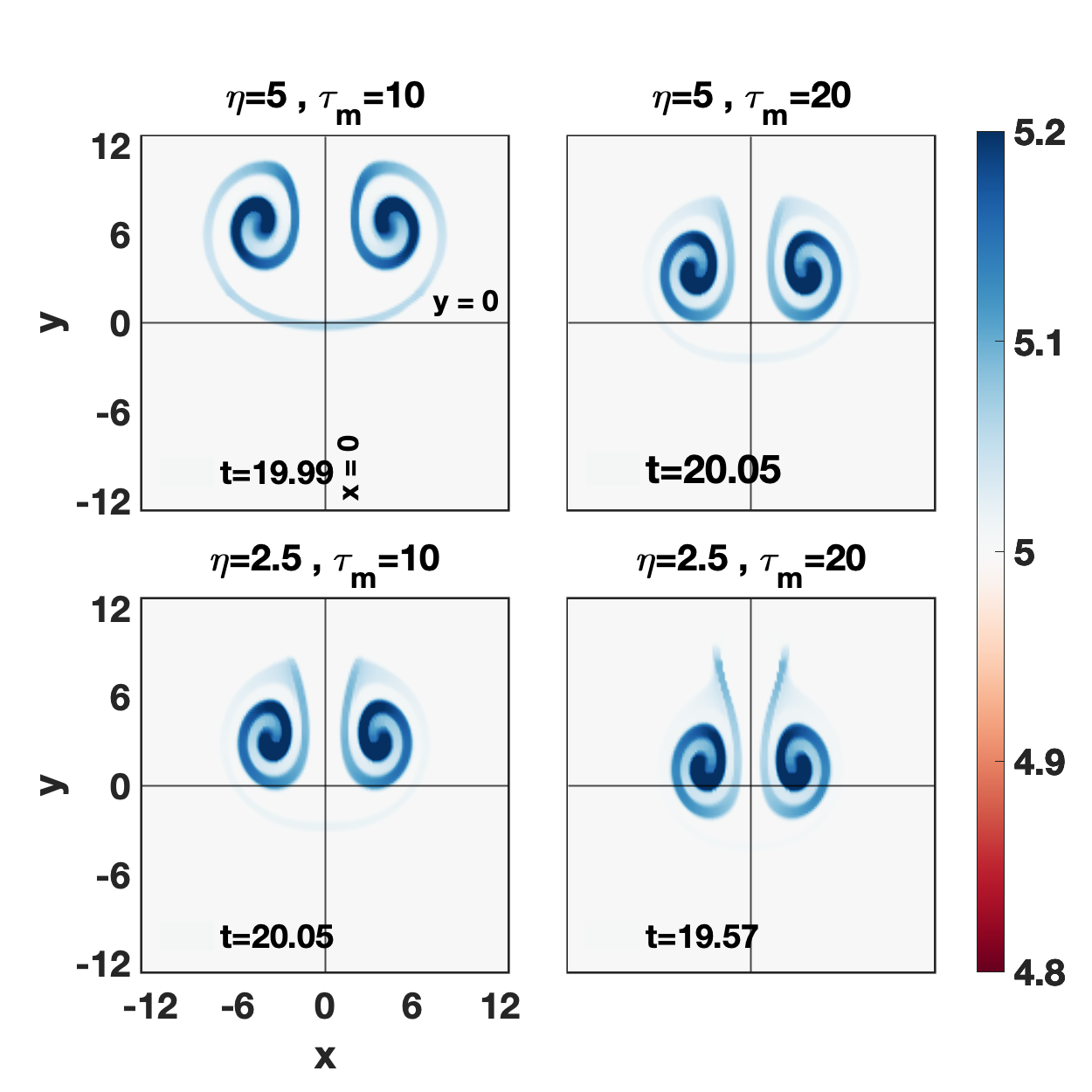}
		\caption{Viscosity $\eta$ vs elasticity $\tau_m$. The snapshots of droplet density  profiles at a particular timestep (see Fig.~\ref{fig:z_vis_elas_vort_drp2by2} for the respective vorticity). The BD instability gets weaker with increasing viscosity (in the vertical direction with a fixed value of $\tau_m$), and elasticity only reduces the viscous effects (along the horizontal direction with a fixed value of $\eta$).}
		\label{fig:z_vis_elas_density_drp2by2}	
	\end{SCfigure}

\begin{SCfigure}[][h]
		\includegraphics[width=0.5\textwidth]{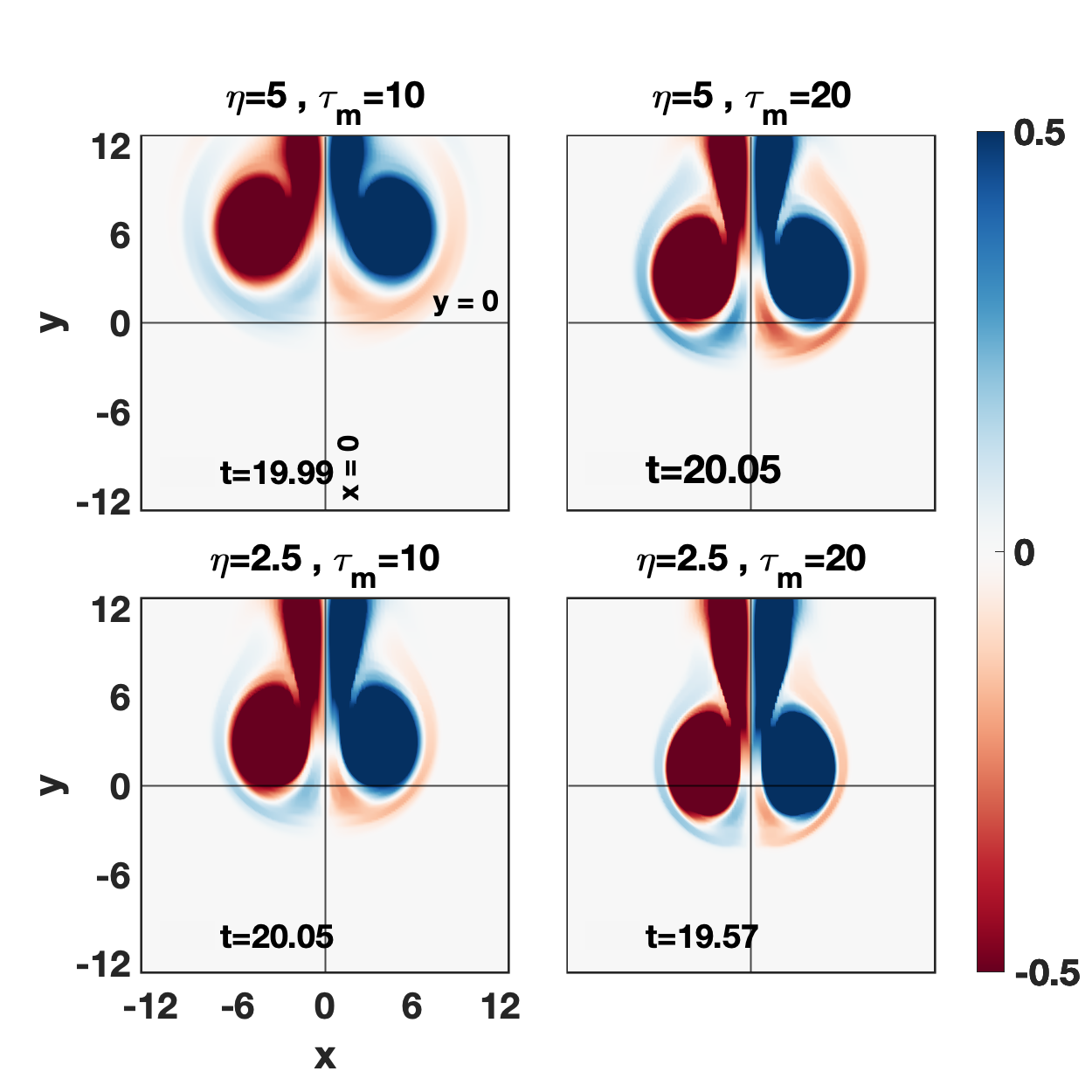} 
		\caption{Viscosity $\eta$ vs elasticity $\tau_m$. The snapshots of droplet vorticity profiles at a particular timestep (see Fig.~\ref{fig:z_vis_elas_density_drp2by2} for the respective density). The BD instability gets weaker with increasing viscosity (in the vertical direction with a fixed value of $\tau_m$), and elasticity only reduces the viscous effects (along the horizontal direction with a fixed value of $\eta$).} 
		\label{fig:z_vis_elas_vort_drp2by2}              
	\end{SCfigure}

	For simulating the falling droplet $i.e$ case B (Fig.~\ref{fig:initialprofb-d}(b)), the values of parameters in Eq.~\ref{eq:rising_bbl} for the droplet density profile  are ${a_c}$=2.0, ${\rho^{\prime}_{0}}=0.5$ and ${x_{c}}=0$, ${y_{c}}=4\pi$. We considered the homogeneous background density as ${\rho_{d0}}=5$. The total density is $\rho_d=\rho_{d0}+\rho_{d1}$. Figure~\ref{fig:drop_den_g10_fluid_b5} shows  the dynamics of the droplet,  falling in tranquil inviscid HD fluid.  The suspended drop starts to fall. As time passes, the drop breaks up forming first a semicircular structure  then a two lobes similar to the case of bubble.
	Figure~\ref{fig:drop_den_g10_ghd_b5_0p35} reveals the different stages of droplet for viscoelsatic fluids.  As done for rising bubble, we shall now compare  Fig.~\ref{fig:drop_den_g10_ghd_b5_0p35}(a) having ${\eta}=3.5, {\tau_m}=10$~$i.e.$~ $\eta/\tau_m$=0.35)   to Fig.~\ref{fig:drop_den_g10_ghd_b5_0p35}(b) having ${\eta}=5, {\tau_m}=10$~$i.e.$~ $\eta/\tau_m$=0.5 . 
	For a  higher coupling strength shown in Fig.~\ref{fig:drop_den_g10_ghd_b5_0p35}(b), the downward motion is slower. The transverse dimension is even larger and the dragging tail does not seem to appear at all. 
	Again, similar to the case of bubbles, the vorticity plots are provided for  observing a clear role of TS waves i.e. pushing the two lobes apart, as seen in Fig.~\ref{fig:drop_vort_g10_all_b5}. 
	For the inviscid HD fluid the corresponding vorticity subplots (Fig.~\ref{fig:drop_vort_g10_all_b5_fluid}) to the mentioned density profile (Fig.~\ref{fig:drop_den_g10_fluid_b5}), no such wave are observed which can constraint the falling rate. Thus, we conclude that the increase of phase velocity of TS waves with coupling strength suppresses the buoyant nature of a droplet. 
	To understand the separate effects of viscosity and elasticity on BD growth the snapshots of density/vorticity snapshots at the same timestep $({\approx}20)$ are shown in Fig.~\ref{fig:z_vis_elas_density_drp2by2}/Fig.~\ref{fig:z_vis_elas_vort_drp2by2}. 
	The role of TS waves can easily be understood from the vorticity snapshots given in Fig. \ref{fig:z_vis_elas_vort_drp2by2}.
	\FloatBarrier
	Similar to the bubble evolution, it can be easily noticed that the BD instability gets weaker with increasing viscosity, and elasticity reduces the viscous effects. Also, in strong coupling limit the expansion of lobes is higher, and disappearance of the dragging tail in density evolution.  Also, the vortex evolution is similar for the same value of coupling strength (${\eta}/{\tau_m}$).  This means the spatio-temporal evolution of  vortices/instabilities depends on the coupling strength.
	\section{Observations and conclusion}
	\label{conclusions}
	We present a detailed discussion on the evolution of Rayleigh-Taylor  instability and  Buoyancy-Driven  instability in two-dimensional strongly coupled dusty plasmas (SCDPs)  in the presence of  gravity. The SCDPs have been treated as viscoelastic fluids in the framework of incompressible generalized hydrodynamics (i- GHD) fluid model. The density  stratification against gravity leading to Rayleigh-Taylor instability. For the buoyancy-driven evolution, we consider a spatially localized  high/low (droplet/bubble) density region placed in a constant  background density medium in the presence of gravity.  Some main observations are as follows:
	\begin{itemize}
		\item {RT instability in the dust fluid phenomenologically similar to the hydrodynamic case as observed in dusty plasma experiment by Pacha {\it et al.} \cite{pacha2012observation}.}
		\item {Our studies, analytically as well as numerical simulations, show that both the instabilities get suppressed  with increasing coupling strength of the medium.}
		\item {The appearance of elasticity $\tau_m$ speed up the growth of RT instability by reducing the effect of viscosity.}
		\item {The spatio-temporal evolution of  vortices/instabilities depends on the value of coupling parameter.}
	\end{itemize}
	The addition of third dimension (3D) will make these results more general. These observations would be presented in a subsequent publication. For the hydrodynamic fluids  Youngs {\it et. al.} reported the difference between 2D and 3D simulation of RT mixing~\cite{youngs1991three} and observed that dissipation of density fluctuations is less in 2D than in 3D. To investigate three-dimensional effects on the Rayleigh–Taylor instability, numerically observe the two-layer roll-up phenomenon of the heavy fluid, which does not occur in the two-dimensional case. \cite{lee2013numerical}
	%
		\input{rt_bd_manuscript.bbl}
\end{document}

%% file: rt_bd_manuscript.bbl
%

%% file: rt_bd_manuscript_vikram.bbl
\begin{thebibliography}{62}%
\makeatletter
\providecommand \@ifxundefined [1]{%
 \@ifx{#1\undefined}
}%
\providecommand \@ifnum [1]{%
 \ifnum #1\expandafter \@firstoftwo
 \else \expandafter \@secondoftwo
 \fi
}%
\providecommand \@ifx [1]{%
 \ifx #1\expandafter \@firstoftwo
 \else \expandafter \@secondoftwo
 \fi
}%
\providecommand \natexlab [1]{#1}%
\providecommand \enquote  [1]{``#1''}%
\providecommand \bibnamefont  [1]{#1}%
\providecommand \bibfnamefont [1]{#1}%
\providecommand \citenamefont [1]{#1}%
\providecommand \href@noop [0]{\@secondoftwo}%
\providecommand \href [0]{\begingroup \@sanitize@url \@href}%
\providecommand \@href[1]{\@@startlink{#1}\@@href}%
\providecommand \@@href[1]{\endgroup#1\@@endlink}%
\providecommand \@sanitize@url [0]{\catcode `\\12\catcode `\$12\catcode
  `\&12\catcode `\#12\catcode `\^12\catcode `\_12\catcode `\%12\relax}%
\providecommand \@@startlink[1]{}%
\providecommand \@@endlink[0]{}%
\providecommand \url  [0]{\begingroup\@sanitize@url \@url }%
\providecommand \@url [1]{\endgroup\@href {#1}{\urlprefix }}%
\providecommand \urlprefix  [0]{URL }%
\providecommand \Eprint [0]{\href }%
\providecommand \doibase [0]{http://dx.doi.org/}%
\providecommand \selectlanguage [0]{\@gobble}%
\providecommand \bibinfo  [0]{\@secondoftwo}%
\providecommand \bibfield  [0]{\@secondoftwo}%
\providecommand \translation [1]{[#1]}%
\providecommand \BibitemOpen [0]{}%
\providecommand \bibitemStop [0]{}%
\providecommand \bibitemNoStop [0]{.\EOS\space}%
\providecommand \EOS [0]{\spacefactor3000\relax}%
\providecommand \BibitemShut  [1]{\csname bibitem#1\endcsname}%
\let\auto@bib@innerbib\@empty
\bibitem [{\citenamefont {Rayleigh}(1900)}]{rayleigh1900investigation}%
  \BibitemOpen
  \bibfield  {author} {\bibinfo {author} {\bibfnamefont {L.}~\bibnamefont
  {Rayleigh}},\ }\bibfield  {title} {\enquote {\bibinfo {title} {Investigation
  of the character of the equilibrium of an incompressible heavy},}\
  }\href@noop {} {\  (\bibinfo {year} {1900})}\BibitemShut {NoStop}%
\bibitem [{\citenamefont {Taylor}(1950)}]{taylor1950instability}%
  \BibitemOpen
  \bibfield  {author} {\bibinfo {author} {\bibfnamefont {G.~I.}\ \bibnamefont
  {Taylor}},\ }\bibfield  {title} {\enquote {\bibinfo {title} {The instability
  of liquid surfaces when accelerated in a direction perpendicular to their
  planes. i},}\ }\href@noop {} {\bibfield  {journal} {\bibinfo  {journal}
  {Proceedings of the Royal Society of London. Series A. Mathematical and
  Physical Sciences}\ }\textbf {\bibinfo {volume} {201}},\ \bibinfo {pages}
  {192--196} (\bibinfo {year} {1950})}\BibitemShut {NoStop}%
\bibitem [{\citenamefont {Chandrasekhar}()}]{chandrhd}%
  \BibitemOpen
  \bibfield  {author} {\bibinfo {author} {\bibfnamefont {S.}~\bibnamefont
  {Chandrasekhar}},\ }\bibfield  {title} {\enquote {\bibinfo {title}
  {Hydrodynamic and hydromagnetic stability, dover, new york, 1981},}\
  }\href@noop {} {\ }\BibitemShut {NoStop}%
\bibitem [{\citenamefont {Youngs}(1984)}]{youngs1984numerical}%
  \BibitemOpen
  \bibfield  {author} {\bibinfo {author} {\bibfnamefont {D.~L.}\ \bibnamefont
  {Youngs}},\ }\bibfield  {title} {\enquote {\bibinfo {title} {Numerical
  simulation of turbulent mixing by rayleigh-taylor instability},}\ }\href@noop
  {} {\bibfield  {journal} {\bibinfo  {journal} {Physica D: Nonlinear
  Phenomena}\ }\textbf {\bibinfo {volume} {12}},\ \bibinfo {pages} {32--44}
  (\bibinfo {year} {1984})}\BibitemShut {NoStop}%
\bibitem [{\citenamefont {Waddell}, \citenamefont {Niederhaus},\ and\
  \citenamefont {Jacobs}(2001)}]{waddell2001experimental}%
  \BibitemOpen
  \bibfield  {author} {\bibinfo {author} {\bibfnamefont {J.}~\bibnamefont
  {Waddell}}, \bibinfo {author} {\bibfnamefont {C.}~\bibnamefont {Niederhaus}},
  \ and\ \bibinfo {author} {\bibfnamefont {J.~W.}\ \bibnamefont {Jacobs}},\
  }\bibfield  {title} {\enquote {\bibinfo {title} {Experimental study of
  rayleigh--taylor instability: low atwood number liquid systems with
  single-mode initial perturbations},}\ }\href@noop {} {\bibfield  {journal}
  {\bibinfo  {journal} {Physics of Fluids}\ }\textbf {\bibinfo {volume} {13}},\
  \bibinfo {pages} {1263--1273} (\bibinfo {year} {2001})}\BibitemShut {NoStop}%
\bibitem [{\citenamefont {Livescu}(2004)}]{livescu2004comp}%
  \BibitemOpen
  \bibfield  {author} {\bibinfo {author} {\bibfnamefont {D.}~\bibnamefont
  {Livescu}},\ }\bibfield  {title} {\enquote {\bibinfo {title} {Compressibility
  effects on the rayleigh--taylor instability growth between immiscible
  fluids},}\ }\href@noop {} {\bibfield  {journal} {\bibinfo  {journal} {Physics
  of fluids}\ }\textbf {\bibinfo {volume} {16}},\ \bibinfo {pages} {118--127}
  (\bibinfo {year} {2004})}\BibitemShut {NoStop}%
\bibitem [{\citenamefont {Guo}\ and\ \citenamefont
  {Tice}(2010)}]{guo2010linear}%
  \BibitemOpen
  \bibfield  {author} {\bibinfo {author} {\bibfnamefont {Y.}~\bibnamefont
  {Guo}}\ and\ \bibinfo {author} {\bibfnamefont {I.}~\bibnamefont {Tice}},\
  }\bibfield  {title} {\enquote {\bibinfo {title} {Linear rayleigh--taylor
  instability for viscous, compressible fluids},}\ }\href@noop {} {\bibfield
  {journal} {\bibinfo  {journal} {SIAM Journal on Mathematical Analysis}\
  }\textbf {\bibinfo {volume} {42}},\ \bibinfo {pages} {1688--1720} (\bibinfo
  {year} {2010})}\BibitemShut {NoStop}%
\bibitem [{\citenamefont {Baldwin}, \citenamefont {Scase},\ and\ \citenamefont
  {Hill}(2015)}]{baldwin2015inhibition}%
  \BibitemOpen
  \bibfield  {author} {\bibinfo {author} {\bibfnamefont {K.~A.}\ \bibnamefont
  {Baldwin}}, \bibinfo {author} {\bibfnamefont {M.~M.}\ \bibnamefont {Scase}},
  \ and\ \bibinfo {author} {\bibfnamefont {R.~J.}\ \bibnamefont {Hill}},\
  }\bibfield  {title} {\enquote {\bibinfo {title} {The inhibition of the
  rayleigh-taylor instability by rotation},}\ }\href@noop {} {\bibfield
  {journal} {\bibinfo  {journal} {Scientific reports}\ }\textbf {\bibinfo
  {volume} {5}},\ \bibinfo {pages} {11706} (\bibinfo {year}
  {2015})}\BibitemShut {NoStop}%
\bibitem [{\citenamefont {Lyubimova}, \citenamefont {Vorobev},\ and\
  \citenamefont {Prokopev}(2019)}]{lyubimova2019rayleigh}%
  \BibitemOpen
  \bibfield  {author} {\bibinfo {author} {\bibfnamefont {T.}~\bibnamefont
  {Lyubimova}}, \bibinfo {author} {\bibfnamefont {A.}~\bibnamefont {Vorobev}},
  \ and\ \bibinfo {author} {\bibfnamefont {S.}~\bibnamefont {Prokopev}},\
  }\bibfield  {title} {\enquote {\bibinfo {title} {Rayleigh-taylor instability
  of a miscible interface in a confined domain},}\ }\href@noop {} {\bibfield
  {journal} {\bibinfo  {journal} {Physics of Fluids}\ }\textbf {\bibinfo
  {volume} {31}},\ \bibinfo {pages} {014104} (\bibinfo {year}
  {2019})}\BibitemShut {NoStop}%
\bibitem [{\citenamefont {Dickinson}\ \emph {et~al.}(1962)\citenamefont
  {Dickinson}, \citenamefont {Bostick}, \citenamefont {DiMarco},\ and\
  \citenamefont {Koslov}}]{dickinson1962experimental}%
  \BibitemOpen
  \bibfield  {author} {\bibinfo {author} {\bibfnamefont {H.}~\bibnamefont
  {Dickinson}}, \bibinfo {author} {\bibfnamefont {W.}~\bibnamefont {Bostick}},
  \bibinfo {author} {\bibfnamefont {J.}~\bibnamefont {DiMarco}}, \ and\
  \bibinfo {author} {\bibfnamefont {S.}~\bibnamefont {Koslov}},\ }\bibfield
  {title} {\enquote {\bibinfo {title} {Experimental study of rayleigh-taylor
  instability in plasma},}\ }\href@noop {} {\bibfield  {journal} {\bibinfo
  {journal} {The Physics of Fluids}\ }\textbf {\bibinfo {volume} {5}},\
  \bibinfo {pages} {1048--1056} (\bibinfo {year} {1962})}\BibitemShut {NoStop}%
\bibitem [{\citenamefont {Ariel}\ and\ \citenamefont
  {Bhatia}(1970)}]{ariel1970rayleigh}%
  \BibitemOpen
  \bibfield  {author} {\bibinfo {author} {\bibfnamefont {P.}~\bibnamefont
  {Ariel}}\ and\ \bibinfo {author} {\bibfnamefont {P.}~\bibnamefont {Bhatia}},\
  }\bibfield  {title} {\enquote {\bibinfo {title} {Rayleigh-taylor instability
  of a rotating plasma},}\ }\href@noop {} {\bibfield  {journal} {\bibinfo
  {journal} {Nuclear Fusion}\ }\textbf {\bibinfo {volume} {10}},\ \bibinfo
  {pages} {141} (\bibinfo {year} {1970})}\BibitemShut {NoStop}%
\bibitem [{\citenamefont {Takabe}\ \emph {et~al.}(1985)\citenamefont {Takabe},
  \citenamefont {Mima}, \citenamefont {Montierth},\ and\ \citenamefont
  {Morse}}]{takabe1985self}%
  \BibitemOpen
  \bibfield  {author} {\bibinfo {author} {\bibfnamefont {H.}~\bibnamefont
  {Takabe}}, \bibinfo {author} {\bibfnamefont {K.}~\bibnamefont {Mima}},
  \bibinfo {author} {\bibfnamefont {L.}~\bibnamefont {Montierth}}, \ and\
  \bibinfo {author} {\bibfnamefont {R.}~\bibnamefont {Morse}},\ }\bibfield
  {title} {\enquote {\bibinfo {title} {Self-consistent growth rate of the
  rayleigh--taylor instability in an ablatively accelerating plasma},}\
  }\href@noop {} {\bibfield  {journal} {\bibinfo  {journal} {The Physics of
  fluids}\ }\textbf {\bibinfo {volume} {28}},\ \bibinfo {pages} {3676--3682}
  (\bibinfo {year} {1985})}\BibitemShut {NoStop}%
\bibitem [{\citenamefont {Mikhailenko}, \citenamefont {Mikhailenko},\ and\
  \citenamefont {Weiland}(2002)}]{mikhailenko2002rayleigh}%
  \BibitemOpen
  \bibfield  {author} {\bibinfo {author} {\bibfnamefont {V.~S.}\ \bibnamefont
  {Mikhailenko}}, \bibinfo {author} {\bibfnamefont {V.~V.}\ \bibnamefont
  {Mikhailenko}}, \ and\ \bibinfo {author} {\bibfnamefont {J.}~\bibnamefont
  {Weiland}},\ }\bibfield  {title} {\enquote {\bibinfo {title}
  {Rayleigh--taylor instability in plasmas with shear flow},}\ }\href@noop {}
  {\bibfield  {journal} {\bibinfo  {journal} {Physics of Plasmas}\ }\textbf
  {\bibinfo {volume} {9}},\ \bibinfo {pages} {2891--2895} (\bibinfo {year}
  {2002})}\BibitemShut {NoStop}%
\bibitem [{\citenamefont {Ma}\ \emph {et~al.}(2006)\citenamefont {Ma},
  \citenamefont {Chen}, \citenamefont {Gan},\ and\ \citenamefont
  {Yu}}]{ma2006behavior}%
  \BibitemOpen
  \bibfield  {author} {\bibinfo {author} {\bibfnamefont {J.}~\bibnamefont
  {Ma}}, \bibinfo {author} {\bibfnamefont {Y.-h.}\ \bibnamefont {Chen}},
  \bibinfo {author} {\bibfnamefont {B.-x.}\ \bibnamefont {Gan}}, \ and\
  \bibinfo {author} {\bibfnamefont {M.}~\bibnamefont {Yu}},\ }\bibfield
  {title} {\enquote {\bibinfo {title} {Behavior of the rayleigh--taylor mode in
  a dusty plasma with rotational and shear flows},}\ }\href@noop {} {\bibfield
  {journal} {\bibinfo  {journal} {Planetary and Space Science}\ }\textbf
  {\bibinfo {volume} {54}},\ \bibinfo {pages} {719--725} (\bibinfo {year}
  {2006})}\BibitemShut {NoStop}%
\bibitem [{\citenamefont {Sen}, \citenamefont {Fukuyama},\ and\ \citenamefont
  {Honary}(2010)}]{sen2010rayleigh}%
  \BibitemOpen
  \bibfield  {author} {\bibinfo {author} {\bibfnamefont {S.}~\bibnamefont
  {Sen}}, \bibinfo {author} {\bibfnamefont {A.}~\bibnamefont {Fukuyama}}, \
  and\ \bibinfo {author} {\bibfnamefont {F.}~\bibnamefont {Honary}},\
  }\bibfield  {title} {\enquote {\bibinfo {title} {Rayleigh taylor instability
  in a dusty plasma},}\ }\href@noop {} {\bibfield  {journal} {\bibinfo
  {journal} {Journal of Atmospheric and Solar-Terrestrial Physics}\ }\textbf
  {\bibinfo {volume} {72}},\ \bibinfo {pages} {938--942} (\bibinfo {year}
  {2010})}\BibitemShut {NoStop}%
\bibitem [{\citenamefont {Weber}\ \emph {et~al.}(2014)\citenamefont {Weber},
  \citenamefont {Clark}, \citenamefont {Cook}, \citenamefont {Busby},\ and\
  \citenamefont {Robey}}]{weber2014inhibition}%
  \BibitemOpen
  \bibfield  {author} {\bibinfo {author} {\bibfnamefont {C.}~\bibnamefont
  {Weber}}, \bibinfo {author} {\bibfnamefont {D.}~\bibnamefont {Clark}},
  \bibinfo {author} {\bibfnamefont {A.}~\bibnamefont {Cook}}, \bibinfo {author}
  {\bibfnamefont {L.}~\bibnamefont {Busby}}, \ and\ \bibinfo {author}
  {\bibfnamefont {H.}~\bibnamefont {Robey}},\ }\bibfield  {title} {\enquote
  {\bibinfo {title} {Inhibition of turbulence in inertial-confinement-fusion
  hot spots by viscous dissipation},}\ }\href@noop {} {\bibfield  {journal}
  {\bibinfo  {journal} {Physical Review E}\ }\textbf {\bibinfo {volume} {89}},\
  \bibinfo {pages} {053106} (\bibinfo {year} {2014})}\BibitemShut {NoStop}%
\bibitem [{\citenamefont {Kessler}, \citenamefont {Dahlburg},\ and\
  \citenamefont {Ganguli}(2014)}]{kessler2014gravitational}%
  \BibitemOpen
  \bibfield  {author} {\bibinfo {author} {\bibfnamefont {D.~A.}\ \bibnamefont
  {Kessler}}, \bibinfo {author} {\bibfnamefont {R.~B.}\ \bibnamefont
  {Dahlburg}}, \ and\ \bibinfo {author} {\bibfnamefont {G.}~\bibnamefont
  {Ganguli}},\ }\bibfield  {title} {\enquote {\bibinfo {title} {Gravitational
  instability and shear stabilization in a dusty plasma layer},}\ }in\
  \href@noop {} {\emph {\bibinfo {booktitle} {2014 United States National
  Committee of URSI National Radio Science Meeting (USNC-URSI NRSM)}}}\
  (\bibinfo {organization} {IEEE},\ \bibinfo {year} {2014})\ pp.\ \bibinfo
  {pages} {1--1}\BibitemShut {NoStop}%
\bibitem [{\citenamefont {Haines}\ \emph {et~al.}(2014)\citenamefont {Haines},
  \citenamefont {Vold}, \citenamefont {Molvig}, \citenamefont {Rauenzahn},\
  and\ \citenamefont {Aldrich}}]{haines2014plasma}%
  \BibitemOpen
  \bibfield  {author} {\bibinfo {author} {\bibfnamefont {B.}~\bibnamefont
  {Haines}}, \bibinfo {author} {\bibfnamefont {E.}~\bibnamefont {Vold}},
  \bibinfo {author} {\bibfnamefont {K.}~\bibnamefont {Molvig}}, \bibinfo
  {author} {\bibfnamefont {R.}~\bibnamefont {Rauenzahn}}, \ and\ \bibinfo
  {author} {\bibfnamefont {C.}~\bibnamefont {Aldrich}},\ }\bibfield  {title}
  {\enquote {\bibinfo {title} {Plasma transport in rayleigh-taylor and
  kelvin-helmholtz instabilities},}\ }\href@noop {} {\bibfield  {journal}
  {\bibinfo  {journal} {Phys. Plasmas}\ }\textbf {\bibinfo {volume} {21}},\
  \bibinfo {pages} {092306} (\bibinfo {year} {2014})}\BibitemShut {NoStop}%
\bibitem [{\citenamefont {Hoshoudy}(2014)}]{hoshoudy2014rayleigh}%
  \BibitemOpen
  \bibfield  {author} {\bibinfo {author} {\bibfnamefont {G.}~\bibnamefont
  {Hoshoudy}},\ }\bibfield  {title} {\enquote {\bibinfo {title}
  {Rayleigh-taylor instability in magnetized plasma},}\ }\href@noop {}
  {\bibfield  {journal} {\bibinfo  {journal} {World Journal of Mechanics}\
  }\textbf {\bibinfo {volume} {4}},\ \bibinfo {pages} {260} (\bibinfo {year}
  {2014})}\BibitemShut {NoStop}%
\bibitem [{\citenamefont {Khiar}\ \emph {et~al.}(2019)\citenamefont {Khiar},
  \citenamefont {Revet}, \citenamefont {Ciardi}, \citenamefont {Burdonov},
  \citenamefont {Filippov}, \citenamefont {B{\'e}ard}, \citenamefont {Cerchez},
  \citenamefont {Chen}, \citenamefont {Gangolf}, \citenamefont {Makarov} \emph
  {et~al.}}]{khiar2019laser}%
  \BibitemOpen
  \bibfield  {author} {\bibinfo {author} {\bibfnamefont {B.}~\bibnamefont
  {Khiar}}, \bibinfo {author} {\bibfnamefont {G.}~\bibnamefont {Revet}},
  \bibinfo {author} {\bibfnamefont {A.}~\bibnamefont {Ciardi}}, \bibinfo
  {author} {\bibfnamefont {K.}~\bibnamefont {Burdonov}}, \bibinfo {author}
  {\bibfnamefont {E.}~\bibnamefont {Filippov}}, \bibinfo {author}
  {\bibfnamefont {J.}~\bibnamefont {B{\'e}ard}}, \bibinfo {author}
  {\bibfnamefont {M.}~\bibnamefont {Cerchez}}, \bibinfo {author} {\bibfnamefont
  {S.}~\bibnamefont {Chen}}, \bibinfo {author} {\bibfnamefont {T.}~\bibnamefont
  {Gangolf}}, \bibinfo {author} {\bibfnamefont {S.}~\bibnamefont {Makarov}},
  \emph {et~al.},\ }\bibfield  {title} {\enquote {\bibinfo {title}
  {Laser-produced magnetic-rayleigh-taylor unstable plasma slabs in a 20 t
  magnetic field},}\ }\href@noop {} {\bibfield  {journal} {\bibinfo  {journal}
  {Physical Review Letters}\ }\textbf {\bibinfo {volume} {123}},\ \bibinfo
  {pages} {205001} (\bibinfo {year} {2019})}\BibitemShut {NoStop}%
\bibitem [{\citenamefont {{Norman}}\ \emph {et~al.}(1981)\citenamefont
  {{Norman}}, \citenamefont {{Smarr}}, \citenamefont {{Smith}},\ and\
  \citenamefont {{Wilson}}}]{Norman_Supernova}%
  \BibitemOpen
  \bibfield  {author} {\bibinfo {author} {\bibfnamefont {M.~L.}\ \bibnamefont
  {{Norman}}}, \bibinfo {author} {\bibfnamefont {L.}~\bibnamefont {{Smarr}}},
  \bibinfo {author} {\bibfnamefont {M.~D.}\ \bibnamefont {{Smith}}}, \ and\
  \bibinfo {author} {\bibfnamefont {J.~R.}\ \bibnamefont {{Wilson}}},\
  }\bibfield  {title} {\enquote {\bibinfo {title} {Hydrodynamic formation of
  twin-exhaust jets},}\ }\href@noop {} {\bibfield  {journal} {\bibinfo
  {journal} {Astrophysical}\ }\textbf {\bibinfo {volume} {247}},\ \bibinfo
  {pages} {52--58} (\bibinfo {year} {1981})}\BibitemShut {NoStop}%
\bibitem [{\citenamefont {Arnett}\ \emph {et~al.}(1989)\citenamefont {Arnett},
  \citenamefont {Bahcall}, \citenamefont {Kirshner},\ and\ \citenamefont
  {Woosley}}]{Arnett_Supernova}%
  \BibitemOpen
  \bibfield  {author} {\bibinfo {author} {\bibfnamefont {W.~D.}\ \bibnamefont
  {Arnett}}, \bibinfo {author} {\bibfnamefont {J.~N.}\ \bibnamefont {Bahcall}},
  \bibinfo {author} {\bibfnamefont {R.~P.}\ \bibnamefont {Kirshner}}, \ and\
  \bibinfo {author} {\bibfnamefont {S.~E.}\ \bibnamefont {Woosley}},\
  }\bibfield  {title} {\enquote {\bibinfo {title} {Supernova 1987a},}\
  }\href@noop {} {\bibfield  {journal} {\bibinfo  {journal} {Annual Review of
  Astronomy and Astrophysics}\ }\textbf {\bibinfo {volume} {27}},\ \bibinfo
  {pages} {629--700} (\bibinfo {year} {1989})}\BibitemShut {NoStop}%
\bibitem [{\citenamefont {Plag}\ and\ \citenamefont
  {J{\"u}ttner}(1995)}]{plag1995rayleigh}%
  \BibitemOpen
  \bibfield  {author} {\bibinfo {author} {\bibfnamefont {H.-P.}\ \bibnamefont
  {Plag}}\ and\ \bibinfo {author} {\bibfnamefont {H.-U.}\ \bibnamefont
  {J{\"u}ttner}},\ }\bibfield  {title} {\enquote {\bibinfo {title}
  {Rayleigh-taylor instabilities of a self-gravitating earth},}\ }\href@noop {}
  {\bibfield  {journal} {\bibinfo  {journal} {Journal of geodynamics}\ }\textbf
  {\bibinfo {volume} {20}},\ \bibinfo {pages} {267--288} (\bibinfo {year}
  {1995})}\BibitemShut {NoStop}%
\bibitem [{\citenamefont {{Allen}}\ and\ \citenamefont
  {{Hughes}}(1984)}]{1984MNRAS}%
  \BibitemOpen
  \bibfield  {author} {\bibinfo {author} {\bibfnamefont {A.~J.}\ \bibnamefont
  {{Allen}}}\ and\ \bibinfo {author} {\bibfnamefont {P.~A.}\ \bibnamefont
  {{Hughes}}},\ }\bibfield  {title} {\enquote {\bibinfo {title} {{The
  Rayleigh-Taylor instability in astrophysical fluids}},}\ }\href@noop {}
  {\bibfield  {journal} {\bibinfo  {journal} {mnras}\ }\textbf {\bibinfo
  {volume} {208}},\ \bibinfo {pages} {609--621} (\bibinfo {year}
  {1984})}\BibitemShut {NoStop}%
\bibitem [{\citenamefont {Zingale}\ \emph {et~al.}(2005)\citenamefont
  {Zingale}, \citenamefont {Woosley}, \citenamefont {Rendleman}, \citenamefont
  {Day},\ and\ \citenamefont {Bell}}]{Zingale_2005}%
  \BibitemOpen
  \bibfield  {author} {\bibinfo {author} {\bibfnamefont {M.}~\bibnamefont
  {Zingale}}, \bibinfo {author} {\bibfnamefont {S.~E.}\ \bibnamefont
  {Woosley}}, \bibinfo {author} {\bibfnamefont {C.~A.}\ \bibnamefont
  {Rendleman}}, \bibinfo {author} {\bibfnamefont {M.~S.}\ \bibnamefont {Day}},
  \ and\ \bibinfo {author} {\bibfnamefont {J.~B.}\ \bibnamefont {Bell}},\
  }\bibfield  {title} {\enquote {\bibinfo {title} {Three-dimensional numerical
  simulations of rayleigh-taylor unstable flames in type ia supernovae},}\
  }\href {\doibase 10.1086/433164} {\bibfield  {journal} {\bibinfo  {journal}
  {The Astrophysical Journal}\ }\textbf {\bibinfo {volume} {632}},\ \bibinfo
  {pages} {1021--1034} (\bibinfo {year} {2005})}\BibitemShut {NoStop}%
\bibitem [{\citenamefont {Keskinen}\ \emph {et~al.}(1981)\citenamefont
  {Keskinen}, \citenamefont {Szuszczewicz}, \citenamefont {Ossakow},\ and\
  \citenamefont {Holmes}}]{keskinen1981nonlinear}%
  \BibitemOpen
  \bibfield  {author} {\bibinfo {author} {\bibfnamefont {M.}~\bibnamefont
  {Keskinen}}, \bibinfo {author} {\bibfnamefont {E.}~\bibnamefont
  {Szuszczewicz}}, \bibinfo {author} {\bibfnamefont {S.}~\bibnamefont
  {Ossakow}}, \ and\ \bibinfo {author} {\bibfnamefont {J.}~\bibnamefont
  {Holmes}},\ }\bibfield  {title} {\enquote {\bibinfo {title} {Nonlinear theory
  and experimental observations of the local collisional rayleigh-taylor
  instability in a descending equatorial spread f ionosphere},}\ }\href@noop {}
  {\bibfield  {journal} {\bibinfo  {journal} {Journal of Geophysical Research:
  Space Physics}\ }\textbf {\bibinfo {volume} {86}},\ \bibinfo {pages}
  {5785--5792} (\bibinfo {year} {1981})}\BibitemShut {NoStop}%
\bibitem [{\citenamefont {Beale}\ and\ \citenamefont
  {Reitz}(1999)}]{Reitz_1999}%
  \BibitemOpen
  \bibfield  {author} {\bibinfo {author} {\bibfnamefont {J.~C.}\ \bibnamefont
  {Beale}}\ and\ \bibinfo {author} {\bibfnamefont {R.~D.}\ \bibnamefont
  {Reitz}},\ }\bibfield  {title} {\enquote {\bibinfo {title} {Modeling spray
  atomization with the kelvin-helmholtz/rayleigh-taylor hybrid model},}\
  }\href@noop {} {\bibfield  {journal} {\bibinfo  {journal} {Atomization and
  sprays}\ }\textbf {\bibinfo {volume} {9}} (\bibinfo {year}
  {1999})}\BibitemShut {NoStop}%
\bibitem [{\citenamefont {Kong}, \citenamefont {Senecal},\ and\ \citenamefont
  {Reitz}(1999)}]{kong1999developments}%
  \BibitemOpen
  \bibfield  {author} {\bibinfo {author} {\bibfnamefont {S.~C.}\ \bibnamefont
  {Kong}}, \bibinfo {author} {\bibfnamefont {P.~K.}\ \bibnamefont {Senecal}}, \
  and\ \bibinfo {author} {\bibfnamefont {R.~D.}\ \bibnamefont {Reitz}},\
  }\bibfield  {title} {\enquote {\bibinfo {title} {Developments in spray
  modeling in diesel and direct-injection gasoline engines},}\ }\href@noop {}
  {\bibfield  {journal} {\bibinfo  {journal} {Oil \& Gas Science and
  Technology}\ }\textbf {\bibinfo {volume} {54}},\ \bibinfo {pages} {197--204}
  (\bibinfo {year} {1999})}\BibitemShut {NoStop}%
\bibitem [{\citenamefont {Chertkov}(2003)}]{Chertkov2003}%
  \BibitemOpen
  \bibfield  {author} {\bibinfo {author} {\bibfnamefont {M.}~\bibnamefont
  {Chertkov}},\ }\bibfield  {title} {\enquote {\bibinfo {title} {Phenomenology
  of rayleigh-taylor turbulence},}\ }\href@noop {} {\bibfield  {journal}
  {\bibinfo  {journal} {Phys. Rev. Lett.}\ }\textbf {\bibinfo {volume} {91}},\
  \bibinfo {pages} {115001} (\bibinfo {year} {2003})}\BibitemShut {NoStop}%
\bibitem [{\citenamefont {Kadau}\ \emph {et~al.}(2004)\citenamefont {Kadau},
  \citenamefont {Germann}, \citenamefont {Hadjiconstantinou}, \citenamefont
  {Lomdahl}, \citenamefont {Dimonte}, \citenamefont {Holian},\ and\
  \citenamefont {Alder}}]{Kadau2004}%
  \BibitemOpen
  \bibfield  {author} {\bibinfo {author} {\bibfnamefont {K.}~\bibnamefont
  {Kadau}}, \bibinfo {author} {\bibfnamefont {T.~C.}\ \bibnamefont {Germann}},
  \bibinfo {author} {\bibfnamefont {N.~G.}\ \bibnamefont {Hadjiconstantinou}},
  \bibinfo {author} {\bibfnamefont {P.~S.}\ \bibnamefont {Lomdahl}}, \bibinfo
  {author} {\bibfnamefont {G.}~\bibnamefont {Dimonte}}, \bibinfo {author}
  {\bibfnamefont {B.~L.}\ \bibnamefont {Holian}}, \ and\ \bibinfo {author}
  {\bibfnamefont {B.~J.}\ \bibnamefont {Alder}},\ }\bibfield  {title} {\enquote
  {\bibinfo {title} {Nanohydrodynamics simulations: An atomistic view of the
  rayleigh-taylor instability},}\ }\href@noop {} {\bibfield  {journal}
  {\bibinfo  {journal} {Proceedings of the National Academy of Sciences of the
  United States of America}\ }\textbf {\bibinfo {volume} {101}},\ \bibinfo
  {pages} {5851--5855} (\bibinfo {year} {2004})}\BibitemShut {NoStop}%
\bibitem [{\citenamefont {Hoshoudy}(2011)}]{hoshoudy2011quantum}%
  \BibitemOpen
  \bibfield  {author} {\bibinfo {author} {\bibfnamefont {G.~A.}\ \bibnamefont
  {Hoshoudy}},\ }\bibfield  {title} {\enquote {\bibinfo {title} {Quantum
  effects on the rayleigh-taylor instability of viscoelastic plasma model
  through a porous medium.}}\ }\href@noop {} {\bibfield  {journal} {\bibinfo
  {journal} {Journal of Modern Physics}\ }\textbf {\bibinfo {volume} {2}}
  (\bibinfo {year} {2011})}\BibitemShut {NoStop}%
\bibitem [{\citenamefont {Srinivasan}\ and\ \citenamefont
  {Tang}(2013)}]{Srinivasan2013}%
  \BibitemOpen
  \bibfield  {author} {\bibinfo {author} {\bibfnamefont {B.}~\bibnamefont
  {Srinivasan}}\ and\ \bibinfo {author} {\bibfnamefont {X.-Z.}\ \bibnamefont
  {Tang}},\ }\bibfield  {title} {\enquote {\bibinfo {title} {The mitigating
  effect of magnetic fields on rayleigh-taylor unstable inertial confinement
  fusion plasmas},}\ }\href@noop {} {\bibfield  {journal} {\bibinfo  {journal}
  {Physics of Plasmas}\ }\textbf {\bibinfo {volume} {20}},\ \bibinfo {eid}
  {056307} (\bibinfo {year} {2013})}\BibitemShut {NoStop}%
\bibitem [{\citenamefont {Debnath}(1994)}]{debnath1994nonlinear}%
  \BibitemOpen
  \bibfield  {author} {\bibinfo {author} {\bibfnamefont {L.}~\bibnamefont
  {Debnath}},\ }\href@noop {} {\emph {\bibinfo {title} {Nonlinear water
  waves}}}\ (\bibinfo  {publisher} {Academic Press},\ \bibinfo {year}
  {1994})\BibitemShut {NoStop}%
\bibitem [{\citenamefont {Abarzhi}(2010)}]{Abarzhi1}%
  \BibitemOpen
  \bibfield  {author} {\bibinfo {author} {\bibfnamefont {S.~I.}\ \bibnamefont
  {Abarzhi}},\ }\bibfield  {title} {\enquote {\bibinfo {title} {Review of
  theoretical modelling approaches of rayleigh--taylor instabilities and
  turbulent mixing},}\ }\href@noop {} {\bibfield  {journal} {\bibinfo
  {journal} {Philosophical Transactions of the Royal Society of London A:
  Mathematical, Physical and Engineering Sciences}\ }\textbf {\bibinfo {volume}
  {368}},\ \bibinfo {pages} {1809--1828} (\bibinfo {year} {2010})}\BibitemShut
  {NoStop}%
\bibitem [{\citenamefont {Abarzhi}(2008)}]{Abarzhi2}%
  \BibitemOpen
  \bibfield  {author} {\bibinfo {author} {\bibfnamefont {S.~I.}\ \bibnamefont
  {Abarzhi}},\ }\bibfield  {title} {\enquote {\bibinfo {title} {Coherent
  structures and pattern formation in the rayleigh-taylor turbulent mixing},}\
  }\href@noop {} {\bibfield  {journal} {\bibinfo  {journal} {Phys. Scr.}\
  }\textbf {\bibinfo {volume} {78}} (\bibinfo {year} {2008})}\BibitemShut
  {NoStop}%
\bibitem [{\citenamefont {Abarzhi}\ and\ \citenamefont
  {Rosner}(2010)}]{Abarzhi3}%
  \BibitemOpen
  \bibfield  {author} {\bibinfo {author} {\bibfnamefont {S.~I.}\ \bibnamefont
  {Abarzhi}}\ and\ \bibinfo {author} {\bibfnamefont {R.}~\bibnamefont
  {Rosner}},\ }\bibfield  {title} {\enquote {\bibinfo {title} {Comparative
  study of approaches for modeling rayleigh-taylor turbulent mixing},}\
  }\href@noop {} {\bibfield  {journal} {\bibinfo  {journal} {Phys. Scr.}\
  }\textbf {\bibinfo {volume} {T142}},\ \bibinfo {pages} {1} (\bibinfo {year}
  {2010})}\BibitemShut {NoStop}%
\bibitem [{\citenamefont {Frenkel}(1955)}]{frenkel_kinetic}%
  \BibitemOpen
  \bibfield  {author} {\bibinfo {author} {\bibfnamefont {J.}~\bibnamefont
  {Frenkel}},\ }\href@noop {} {\emph {\bibinfo {title} {Kinetic Theory Of
  Liquids}}}\ (\bibinfo  {publisher} {Dover Publications},\ \bibinfo {year}
  {1955})\BibitemShut {NoStop}%
\bibitem [{\citenamefont {Kaus}\ and\ \citenamefont
  {Becker}(2007)}]{kaus2007effects}%
  \BibitemOpen
  \bibfield  {author} {\bibinfo {author} {\bibfnamefont {B.~J.}\ \bibnamefont
  {Kaus}}\ and\ \bibinfo {author} {\bibfnamefont {T.~W.}\ \bibnamefont
  {Becker}},\ }\bibfield  {title} {\enquote {\bibinfo {title} {Effects of
  elasticity on the rayleigh--taylor instability: implications for large-scale
  geodynamics},}\ }\href@noop {} {\bibfield  {journal} {\bibinfo  {journal}
  {Geophysical Journal International}\ }\textbf {\bibinfo {volume} {168}},\
  \bibinfo {pages} {843--862} (\bibinfo {year} {2007})}\BibitemShut {NoStop}%
\bibitem [{\citenamefont {Guido}\ \emph {et~al.}(2010)\citenamefont {Guido},
  \citenamefont {Mazzino}, \citenamefont {Musacchio},\ and\ \citenamefont
  {Vozella}}]{boffetta_mazzino_musacchio_vozella_2010}%
  \BibitemOpen
  \bibfield  {author} {\bibinfo {author} {\bibfnamefont {B.}~\bibnamefont
  {Guido}}, \bibinfo {author} {\bibfnamefont {A.}~\bibnamefont {Mazzino}},
  \bibinfo {author} {\bibfnamefont {S.}~\bibnamefont {Musacchio}}, \ and\
  \bibinfo {author} {\bibfnamefont {L.}~\bibnamefont {Vozella}},\ }\bibfield
  {title} {\enquote {\bibinfo {title} {Rayleigh–taylor instability in a
  viscoelastic binary fluid},}\ }\href {\doibase 10.1017/S0022112009992497}
  {\bibfield  {journal} {\bibinfo  {journal} {Journal of Fluid Mechanics}\
  }\textbf {\bibinfo {volume} {643}},\ \bibinfo {pages} {127–136} (\bibinfo
  {year} {2010})}\BibitemShut {NoStop}%
\bibitem [{\citenamefont {Kaw}\ and\ \citenamefont {Sen}(1998)}]{Kaw_Sen_1998}%
  \BibitemOpen
  \bibfield  {author} {\bibinfo {author} {\bibfnamefont {P.~K.}\ \bibnamefont
  {Kaw}}\ and\ \bibinfo {author} {\bibfnamefont {A.}~\bibnamefont {Sen}},\
  }\bibfield  {title} {\enquote {\bibinfo {title} {Low frequency modes in
  strongly coupled dusty plasmas},}\ }\href@noop {} {\bibfield  {journal}
  {\bibinfo  {journal} {Physics of Plasmas}\ }\textbf {\bibinfo {volume} {5}},\
  \bibinfo {pages} {3552--3559} (\bibinfo {year} {1998})}\BibitemShut {NoStop}%
\bibitem [{\citenamefont {Kaw}(2001)}]{Kaw_2001}%
  \BibitemOpen
  \bibfield  {author} {\bibinfo {author} {\bibfnamefont {P.~K.}\ \bibnamefont
  {Kaw}},\ }\bibfield  {title} {\enquote {\bibinfo {title} {Collective modes in
  a strongly coupled dusty plasma},}\ }\href@noop {} {\bibfield  {journal}
  {\bibinfo  {journal} {Physics of Plasmas}\ }\textbf {\bibinfo {volume} {8}},\
  \bibinfo {pages} {1870--1878} (\bibinfo {year} {2001})}\BibitemShut {NoStop}%
\bibitem [{\citenamefont {Tiwari}\ \emph {et~al.}(2012)\citenamefont {Tiwari},
  \citenamefont {Das}, \citenamefont {Angom}, \citenamefont {Patel},\ and\
  \citenamefont {Kaw}}]{tiwari2012kelvin}%
  \BibitemOpen
  \bibfield  {author} {\bibinfo {author} {\bibfnamefont {S.~K.}\ \bibnamefont
  {Tiwari}}, \bibinfo {author} {\bibfnamefont {A.}~\bibnamefont {Das}},
  \bibinfo {author} {\bibfnamefont {D.}~\bibnamefont {Angom}}, \bibinfo
  {author} {\bibfnamefont {B.~G.}\ \bibnamefont {Patel}}, \ and\ \bibinfo
  {author} {\bibfnamefont {P.}~\bibnamefont {Kaw}},\ }\bibfield  {title}
  {\enquote {\bibinfo {title} {Kelvin-helmholtz instability in a strongly
  coupled dusty plasma medium},}\ }\href@noop {} {\bibfield  {journal}
  {\bibinfo  {journal} {Physics of Plasmas}\ }\textbf {\bibinfo {volume}
  {19}},\ \bibinfo {pages} {073703} (\bibinfo {year} {2012})}\BibitemShut
  {NoStop}%
\bibitem [{\citenamefont {S.~Dharodi}, \citenamefont {K.~Tiwari},\ and\
  \citenamefont {Das}(2014)}]{dharodi2014visco}%
  \BibitemOpen
  \bibfield  {author} {\bibinfo {author} {\bibfnamefont {V.}~\bibnamefont
  {S.~Dharodi}}, \bibinfo {author} {\bibfnamefont {S.}~\bibnamefont
  {K.~Tiwari}}, \ and\ \bibinfo {author} {\bibfnamefont {A.}~\bibnamefont
  {Das}},\ }\bibfield  {title} {\enquote {\bibinfo {title} {Visco-elastic fluid
  simulations of coherent structures in strongly coupled dusty plasma
  medium},}\ }\href@noop {} {\bibfield  {journal} {\bibinfo  {journal} {Physics
  of Plasmas}\ }\textbf {\bibinfo {volume} {21}},\ \bibinfo {pages} {073705}
  (\bibinfo {year} {2014})}\BibitemShut {NoStop}%
\bibitem [{\citenamefont {Tiwari}\ \emph {et~al.}(2014)\citenamefont {Tiwari},
  \citenamefont {Dharodi}, \citenamefont {Das}, \citenamefont {Patel},\ and\
  \citenamefont {Kaw}}]{tiwari2014evolution}%
  \BibitemOpen
  \bibfield  {author} {\bibinfo {author} {\bibfnamefont {S.~K.}\ \bibnamefont
  {Tiwari}}, \bibinfo {author} {\bibfnamefont {V.~S.}\ \bibnamefont {Dharodi}},
  \bibinfo {author} {\bibfnamefont {A.}~\bibnamefont {Das}}, \bibinfo {author}
  {\bibfnamefont {B.~G.}\ \bibnamefont {Patel}}, \ and\ \bibinfo {author}
  {\bibfnamefont {P.}~\bibnamefont {Kaw}},\ }\bibfield  {title} {\enquote
  {\bibinfo {title} {Evolution of sheared flow structure in visco-elastic
  fluids},}\ }\href@noop {} {\bibfield  {journal} {\bibinfo  {journal} {AIP
  Conf. Proc}\ }\textbf {\bibinfo {volume} {1582}},\ \bibinfo {pages} {55}
  (\bibinfo {year} {2014})}\BibitemShut {NoStop}%
\bibitem [{\citenamefont {Dharodi}\ \emph {et~al.}(2016)\citenamefont
  {Dharodi}, \citenamefont {Das}, \citenamefont {Patel},\ and\ \citenamefont
  {Kaw}}]{dharodi2016sub}%
  \BibitemOpen
  \bibfield  {author} {\bibinfo {author} {\bibfnamefont {V.~S.}\ \bibnamefont
  {Dharodi}}, \bibinfo {author} {\bibfnamefont {A.}~\bibnamefont {Das}},
  \bibinfo {author} {\bibfnamefont {B.~G.}\ \bibnamefont {Patel}}, \ and\
  \bibinfo {author} {\bibfnamefont {P.~K.}\ \bibnamefont {Kaw}},\ }\bibfield
  {title} {\enquote {\bibinfo {title} {Sub-and super-luminar propagation of
  structures satisfying poynting-like theorem for incompressible generalized
  hydrodynamic fluid model depicting strongly coupled dusty plasma medium},}\
  }\href@noop {} {\bibfield  {journal} {\bibinfo  {journal} {Physics of
  Plasmas}\ }\textbf {\bibinfo {volume} {23}},\ \bibinfo {pages} {013707}
  (\bibinfo {year} {2016})}\BibitemShut {NoStop}%
\bibitem [{\citenamefont {Das}\ and\ \citenamefont {Kaw}(2014)}]{Das_2014}%
  \BibitemOpen
  \bibfield  {author} {\bibinfo {author} {\bibfnamefont {A.}~\bibnamefont
  {Das}}\ and\ \bibinfo {author} {\bibfnamefont {P.~K.}\ \bibnamefont {Kaw}},\
  }\bibfield  {title} {\enquote {\bibinfo {title} {Suppression of
  rayleigh-taylor instability in strongly coupled plasmas},}\ }\href@noop {}
  {\bibfield  {journal} {\bibinfo  {journal} {Physics of Plasmas}\ }\textbf
  {\bibinfo {volume} {21}} (\bibinfo {year} {2014})}\BibitemShut {NoStop}%
\bibitem [{\citenamefont {Dharodi}(2020)}]{dharodi2020rotating}%
  \BibitemOpen
  \bibfield  {author} {\bibinfo {author} {\bibfnamefont {V.~S.}\ \bibnamefont
  {Dharodi}},\ }\bibfield  {title} {\enquote {\bibinfo {title} {Rotating
  vortices in two-dimensional inhomogeneous strongly coupled dusty plasmas:
  shear and spiral-density waves},}\ }\href@noop {} {\bibfield  {journal}
  {\bibinfo  {journal} {arXiv preprint arXiv:2006.04545}\ } (\bibinfo {year}
  {2020})}\BibitemShut {NoStop}%
\bibitem [{\citenamefont {Biot}(1965)}]{biot1965mechanics}%
  \BibitemOpen
  \bibfield  {author} {\bibinfo {author} {\bibfnamefont {M.~A.}\ \bibnamefont
  {Biot}},\ }\href@noop {} {\emph {\bibinfo {title} {Mechanics of incremental
  deformations}}}\ (\bibinfo {year} {1965})\BibitemShut {NoStop}%
\bibitem [{\citenamefont {Biot}\ and\ \citenamefont
  {Od{\'e}}(1965)}]{biot1965theory}%
  \BibitemOpen
  \bibfield  {author} {\bibinfo {author} {\bibfnamefont {M.~A.}\ \bibnamefont
  {Biot}}\ and\ \bibinfo {author} {\bibfnamefont {H.}~\bibnamefont {Od{\'e}}},\
  }\bibfield  {title} {\enquote {\bibinfo {title} {Theory of gravity
  instability with variable overburden and compaction},}\ }\href@noop {}
  {\bibfield  {journal} {\bibinfo  {journal} {Geophysics}\ }\textbf {\bibinfo
  {volume} {30}},\ \bibinfo {pages} {213--227} (\bibinfo {year}
  {1965})}\BibitemShut {NoStop}%
\bibitem [{\citenamefont {Od{\'e}}(1966)}]{ode1966gravitational}%
  \BibitemOpen
  \bibfield  {author} {\bibinfo {author} {\bibfnamefont {H.}~\bibnamefont
  {Od{\'e}}},\ }\bibfield  {title} {\enquote {\bibinfo {title} {Gravitational
  instability of a multilayered system of high viscosity: Verhandelingen der
  koninklijke nederlandse akademie van wetenschappen, afdeeling natuurkunde:
  Reeks 1, wiskunde, natuurkunde, scheikunde, aardkunde, technische
  wetenschappen},}\ }\href@noop {} {\  (\bibinfo {year} {1966})}\BibitemShut
  {NoStop}%
\bibitem [{\citenamefont {Poliakov}\ \emph {et~al.}(1993)\citenamefont
  {Poliakov}, \citenamefont {Cundall}, \citenamefont {Podladchikov},\ and\
  \citenamefont {Lyakhovsky}}]{poliakov1993explicit}%
  \BibitemOpen
  \bibfield  {author} {\bibinfo {author} {\bibfnamefont {A.}~\bibnamefont
  {Poliakov}}, \bibinfo {author} {\bibfnamefont {P.}~\bibnamefont {Cundall}},
  \bibinfo {author} {\bibfnamefont {Y.}~\bibnamefont {Podladchikov}}, \ and\
  \bibinfo {author} {\bibfnamefont {V.}~\bibnamefont {Lyakhovsky}},\ }\bibfield
   {title} {\enquote {\bibinfo {title} {An explicit inertial method for the
  simulation of viscoelastic flow: an evaluation of elastic effects on diapiric
  flow in two-and three-layers models},}\ }in\ \href@noop {} {\emph {\bibinfo
  {booktitle} {Flow and Creep in the Solar System: observations, modeling and
  Theory}}}\ (\bibinfo  {publisher} {Springer},\ \bibinfo {year} {1993})\ pp.\
  \bibinfo {pages} {175--195}\BibitemShut {NoStop}%
\bibitem [{\citenamefont {Naimark}\ and\ \citenamefont
  {Ismail-Zadeh}(1994)}]{naimark1994gravitational}%
  \BibitemOpen
  \bibfield  {author} {\bibinfo {author} {\bibfnamefont {B.}~\bibnamefont
  {Naimark}}\ and\ \bibinfo {author} {\bibfnamefont {A.}~\bibnamefont
  {Ismail-Zadeh}},\ }\bibfield  {title} {\enquote {\bibinfo {title}
  {Gravitational instability of maxwell upper mantle},}\ }\href@noop {}
  {\bibfield  {journal} {\bibinfo  {journal} {Comput. Seis. Geodyn}\ }\textbf
  {\bibinfo {volume} {1}},\ \bibinfo {pages} {36--42} (\bibinfo {year}
  {1994})}\BibitemShut {NoStop}%
\bibitem [{\citenamefont {Avinash}\ and\ \citenamefont
  {Sen}(2015)}]{avinash2015rayleigh}%
  \BibitemOpen
  \bibfield  {author} {\bibinfo {author} {\bibfnamefont {K.}~\bibnamefont
  {Avinash}}\ and\ \bibinfo {author} {\bibfnamefont {A.}~\bibnamefont {Sen}},\
  }\bibfield  {title} {\enquote {\bibinfo {title} {Rayleigh-taylor instability
  in dusty plasma experiment},}\ }\href@noop {} {\bibfield  {journal} {\bibinfo
   {journal} {Physics of Plasmas}\ }\textbf {\bibinfo {volume} {22}},\ \bibinfo
  {pages} {083707} (\bibinfo {year} {2015})}\BibitemShut {NoStop}%
\bibitem [{\citenamefont {Dharodi}\ and\ \citenamefont
  {Das}(2014)}]{rt_dharodi2014}%
  \BibitemOpen
  \bibfield  {author} {\bibinfo {author} {\bibfnamefont {V.~S.}\ \bibnamefont
  {Dharodi}}\ and\ \bibinfo {author} {\bibfnamefont {A.}~\bibnamefont {Das}},\
  }\href@noop {} {\enquote {\bibinfo {title} {Rayleigh-taylor instability in a
  visco-elastic medium using generalized hydrodynamic model},}\ } (\bibinfo
  {year} {15-19 September 2014}),\ \bibinfo {note} {17th international congress
  on plasma physics at Instituto Superior Tecnico, Lisbon,
  Portugal}\BibitemShut {NoStop}%
\bibitem [{\citenamefont {Boris}\ \emph {et~al.}(1993)\citenamefont {Boris},
  \citenamefont {Landsberg}, \citenamefont {Oran},\ and\ \citenamefont
  {Gardner}}]{boris_book}%
  \BibitemOpen
  \bibfield  {author} {\bibinfo {author} {\bibfnamefont {J.~P.}\ \bibnamefont
  {Boris}}, \bibinfo {author} {\bibfnamefont {A.~M.}\ \bibnamefont
  {Landsberg}}, \bibinfo {author} {\bibfnamefont {E.~S.}\ \bibnamefont {Oran}},
  \ and\ \bibinfo {author} {\bibfnamefont {J.~H.}\ \bibnamefont {Gardner}},\
  }\href@noop {} {\emph {\bibinfo {title} {LCPFCT A flux-corrected transport
  algorithm for solving generalized continuity equations}}}\ (\bibinfo
  {publisher} {Technical Report NRL Memorandum Report 93-7192, Naval Research
  Laboratory},\ \bibinfo {year} {1993})\BibitemShut {NoStop}%
\bibitem [{\citenamefont {Swarztrauber}, \citenamefont {Sweet},\ and\
  \citenamefont {Adams}(1999)}]{swarztrauber1999fishpack}%
  \BibitemOpen
  \bibfield  {author} {\bibinfo {author} {\bibfnamefont {P.}~\bibnamefont
  {Swarztrauber}}, \bibinfo {author} {\bibfnamefont {R.}~\bibnamefont {Sweet}},
  \ and\ \bibinfo {author} {\bibfnamefont {J.~C.}\ \bibnamefont {Adams}},\
  }\bibfield  {title} {\enquote {\bibinfo {title} {Fishpack: Efficient fortran
  subprograms for the solution of elliptic partial differential equations},}\
  }\href@noop {} {\bibfield  {journal} {\bibinfo  {journal} {UCAR Publication,
  July}\ } (\bibinfo {year} {1999})}\BibitemShut {NoStop}%
\bibitem [{\citenamefont {Hosseini}\ \emph {et~al.}(2017)\citenamefont
  {Hosseini}, \citenamefont {Turek}, \citenamefont {M{\"o}ller},\ and\
  \citenamefont {Palmes}}]{hosseini2017isogeometric}%
  \BibitemOpen
  \bibfield  {author} {\bibinfo {author} {\bibfnamefont {B.~S.}\ \bibnamefont
  {Hosseini}}, \bibinfo {author} {\bibfnamefont {S.}~\bibnamefont {Turek}},
  \bibinfo {author} {\bibfnamefont {M.}~\bibnamefont {M{\"o}ller}}, \ and\
  \bibinfo {author} {\bibfnamefont {C.}~\bibnamefont {Palmes}},\ }\bibfield
  {title} {\enquote {\bibinfo {title} {Isogeometric analysis of the
  navier--stokes--cahn--hilliard equations with application to incompressible
  two-phase flows},}\ }\href@noop {} {\bibfield  {journal} {\bibinfo  {journal}
  {Journal of Computational Physics}\ }\textbf {\bibinfo {volume} {348}},\
  \bibinfo {pages} {171--194} (\bibinfo {year} {2017})}\BibitemShut {NoStop}%
\bibitem [{\citenamefont {Talat}\ \emph {et~al.}(2018)\citenamefont {Talat},
  \citenamefont {Mavri{\v{c}}}, \citenamefont {Hati{\'c}}, \citenamefont
  {Bajt},\ and\ \citenamefont {{\v{S}}arler}}]{talat2018phase}%
  \BibitemOpen
  \bibfield  {author} {\bibinfo {author} {\bibfnamefont {N.}~\bibnamefont
  {Talat}}, \bibinfo {author} {\bibfnamefont {B.}~\bibnamefont {Mavri{\v{c}}}},
  \bibinfo {author} {\bibfnamefont {V.}~\bibnamefont {Hati{\'c}}}, \bibinfo
  {author} {\bibfnamefont {S.}~\bibnamefont {Bajt}}, \ and\ \bibinfo {author}
  {\bibfnamefont {B.}~\bibnamefont {{\v{S}}arler}},\ }\bibfield  {title}
  {\enquote {\bibinfo {title} {Phase field simulation of rayleigh--taylor
  instability with a meshless method},}\ }\href@noop {} {\bibfield  {journal}
  {\bibinfo  {journal} {Engineering Analysis with Boundary Elements}\ }\textbf
  {\bibinfo {volume} {87}},\ \bibinfo {pages} {78--89} (\bibinfo {year}
  {2018})}\BibitemShut {NoStop}%
\bibitem [{\citenamefont {Pacha}\ \emph {et~al.}(2012)\citenamefont {Pacha},
  \citenamefont {Heinrich}, \citenamefont {Kim},\ and\ \citenamefont
  {Merlino}}]{pacha2012observation}%
  \BibitemOpen
  \bibfield  {author} {\bibinfo {author} {\bibfnamefont {K.~A.}\ \bibnamefont
  {Pacha}}, \bibinfo {author} {\bibfnamefont {J.~R.}\ \bibnamefont {Heinrich}},
  \bibinfo {author} {\bibfnamefont {S.-H.}\ \bibnamefont {Kim}}, \ and\
  \bibinfo {author} {\bibfnamefont {R.~L.}\ \bibnamefont {Merlino}},\
  }\bibfield  {title} {\enquote {\bibinfo {title} {Observation of the taylor
  instability in a dusty plasma},}\ }\href@noop {} {\bibfield  {journal}
  {\bibinfo  {journal} {Physics of Plasmas}\ }\textbf {\bibinfo {volume}
  {19}},\ \bibinfo {pages} {014501} (\bibinfo {year} {2012})}\BibitemShut
  {NoStop}%
\bibitem [{\citenamefont {Horstmann}\ \emph {et~al.}(2014)\citenamefont
  {Horstmann}, \citenamefont {Henningsson}, \citenamefont {Thomas},\ and\
  \citenamefont {Bomphrey}}]{horstmann2014wake}%
  \BibitemOpen
  \bibfield  {author} {\bibinfo {author} {\bibfnamefont {J.~T.}\ \bibnamefont
  {Horstmann}}, \bibinfo {author} {\bibfnamefont {P.}~\bibnamefont
  {Henningsson}}, \bibinfo {author} {\bibfnamefont {A.~L.}\ \bibnamefont
  {Thomas}}, \ and\ \bibinfo {author} {\bibfnamefont {R.~J.}\ \bibnamefont
  {Bomphrey}},\ }\bibfield  {title} {\enquote {\bibinfo {title} {Wake
  development behind paired wings with tip and root trailing vortices:
  consequences for animal flight force estimates},}\ }\href@noop {} {\bibfield
  {journal} {\bibinfo  {journal} {PloS one}\ }\textbf {\bibinfo {volume} {9}}
  (\bibinfo {year} {2014})}\BibitemShut {NoStop}%
\bibitem [{\citenamefont {Youngs}(1991)}]{youngs1991three}%
  \BibitemOpen
  \bibfield  {author} {\bibinfo {author} {\bibfnamefont {D.~L.}\ \bibnamefont
  {Youngs}},\ }\bibfield  {title} {\enquote {\bibinfo {title}
  {Three-dimensional numerical simulation of turbulent mixing by
  rayleigh--taylor instability},}\ }\href@noop {} {\bibfield  {journal}
  {\bibinfo  {journal} {Physics of Fluids A: Fluid Dynamics}\ }\textbf
  {\bibinfo {volume} {3}},\ \bibinfo {pages} {1312--1320} (\bibinfo {year}
  {1991})}\BibitemShut {NoStop}%
\bibitem [{\citenamefont {Lee}\ and\ \citenamefont
  {Kim}(2013)}]{lee2013numerical}%
  \BibitemOpen
  \bibfield  {author} {\bibinfo {author} {\bibfnamefont {H.~G.}\ \bibnamefont
  {Lee}}\ and\ \bibinfo {author} {\bibfnamefont {J.}~\bibnamefont {Kim}},\
  }\bibfield  {title} {\enquote {\bibinfo {title} {Numerical simulation of the
  three-dimensional rayleigh--taylor instability},}\ }\href@noop {} {\bibfield
  {journal} {\bibinfo  {journal} {Computers \& Mathematics with Applications}\
  }\textbf {\bibinfo {volume} {66}},\ \bibinfo {pages} {1466--1474} (\bibinfo
  {year} {2013})}\BibitemShut {NoStop}%
\end{thebibliography}
